# Chern number reversal and emergent superconductivity in rhombohedral graphene induced by in-plane magnetic fields

Xiaozhou Zan[1,10]*, Hangzhe Li[1,10], Jiawei Guo[1,10], Gengdong Zhou[2,10], Kangyao Chen[1], Cihan Gao[1], Zijun Xu[1], Kenji Watanabe[3], Takashi Taniguchi[4], Anqi Wang[5], Jie Shen[5,6], Jinsong Zhang[1,7,9], Zhida Song[2,7,8]* and Yayu Wang[1,7,9]*

[1]*State Key Laboratory of Low Dimensional Quantum Physics, Department of Physics, Tsinghua University, Beijing 100084, China.*

[2]*International Center for Quantum Materials, School of Physics, Peking University, Beijing 100871, China.*

[3]*Research Center for Electronic and Optical Materials, National Institute for Materials Science, 1-1 Namiki, Tsukuba 305-0044, Japan.*

[4]*Research Center for Materials Nanoarchitectonics, National Institute for Materials Science, 1-1 Namiki, Tsukuba 305-0044, Japan.*

[5]*Beijing National Laboratory for Condensed Matter Physics, Institute of Physics, Chinese Academy of Sciences, Beijing 100190, China.*

[6]*School of Physical Sciences, University of Chinese Academy of Sciences, Beijing 100190, China.*

[7]*Hefei National Laboratory, Hefei 230088, China.*

[8]*Beijing Key Laboratory of Quantum Devices, Peking University, Beijing 100871, China.*

[9]*New Cornerstone Science Laboratory, Frontier Science Center for Quantum Information, Beijing 100084, China.*

[10]*These authors contributed equally: Xiaozhou Zan, Hangzhe Li, Jiawei Guo, Gengdong Zhou*

*Corresponding authors: *zanxz@tsinghua.edu.cn; songzd@pku.edu.cn; yayuwang@tsinghua.edu.cn*

**Rhombohedral graphene with topological flat bands offers an ideal platform for realizing correlated and topological quantum phases. Here we investigate hBN-aligned eight-layer rhombohedral graphene moiré superlattices, which host a robust quantum anomalous Hall (QAH) state alongside three unconventional superconducting phases. For electron-doped carriers away from the moiré potential, we observe QAH Chern number reversal driven by the displacement fields and in-plane magnetic fields ($B_\parallel$). For hole-doped carriers near the moiré superlattice, the three superconducting phases exhibit distinctively different $B_\parallel$ responses: one is weakly enhanced, the second is strongly suppressed, and the third exclusively induced by $B_\parallel$. The isotropic $B_\parallel$ response in the QAH regime points to interplay between orbital magnetism and spin-orbit coupling, and the field-emergent superconductivity provides compelling evidence for spin-triplet pairing. Our work demonstrates a highly versatile platform for coexisting topological and superconducting states, and highlights $B_\parallel$ as a powerful in-situ control knob for engineering novel quantum devices.**

In two-dimensional (2D) moiré systems, flat bands with net Berry curvature provide an ideal platform for realizing novel correlated and topological phases of matter (*1-15*). A particularly versatile system is the moiré superlattice formed by rhombohedral graphene aligned with hexagonal boron nitride (hBN), which has been found to host correlated insulating states, integer and fractional quantum anomalous Hall (QAH) states, and unconventional superconducting (SC) states (*16-40*). The 2D confinement in graphene-based systems restricts electron motion within the plane, giving rise to a perpendicular orbital magnetic moment (*28,41*). Consequently, out-of-plane magnetic fields ($B_\perp$) have been regarded as the primary tuning knob for orbital magnetism and valley polarization, whereas in-plane magnetic fields ($B_\parallel$) are expected to exert comparatively minor effects. There are recent studies on the role of $B_\parallel$ in thin rhombohedral graphene moiré superlattices (*42-43*), but the impact of coexisting $B_\perp$ and $B_\parallel$ on topological and SC states in thicker devices near the bulk regime remains largely unexplored.

With increasing layer number in rhombohedral graphene, the surface-localized electronic wavefunctions become more pronounced and the band structure becomes more complex, exhibiting enhanced flat-band characteristics and larger Berry curvature (*44-50*). These trends amplify the roles of Coulomb interaction, orbital magnetism, and band topology (*51-53*), making thicker rhombohedral graphene moiré superlattices especially promising for the emergence of intertwined topological phases, SC states, and symmetry-breaking orders (*54-59*). However, the fabrication of high-quality, pure-phase rhombohedral graphene moiré superlattices approaching the bulk regime poses substantial challenges, owing to the extremely rapid proliferation of stacking orders with increasing layer number.

Here, we report the observation of a robust QAH state with electron-doped carriers away from the moiré superlattice, and three unconventional SC phases with hole-doped carriers near the moiré superlattice in an 8-layer rhombohedral graphene (8L-RG) moiré superlattice aligned with hBN. The Chern number and coercivity of the QAH states can be systematically controlled by $B_\parallel$ in an isotropic manner, and the three SC phases exhibit distinctively different responses to $B_\parallel$. These results reveal the critical role of the spin degree of freedom in the correlated and topological phases in rhombohedral graphene, and the flexible in-situ tuning of QAH and SC phases by $B_\parallel$ makes this system uniquely suitable for exploring more exotic phenomena and device applications. Measurements on two devices with similar moiré periods ~ 12.7 nm (twist angle ~ 0.55°) exhibit consistent results. The data presented in the main text are from Device D1, and the data from Device D2 are documented in Supplementary Note 9 and fig. S13.

**Phase diagram of the QAH state**

Fig. 1A shows a schematic of the 8L-RG moiré superlattice device. The dual-gate architecture enables independent tuning of the displacement field $D$ and carrier density $n$, defined as $D = (C_b V_{bg} - C_t V_{tg})/2\varepsilon_0$ and $n = (C_b V_{bg} + C_t V_{tg})/e$, where $e$ is the elementary charge, $\varepsilon_0$ is the vacuum permittivity, and $C_b$ ($C_t$) and $V_{bg}$ ($V_{tg}$) denote the geometrical capacitance per unit area and the applied voltage of the bottom (top) gate, respectively. Further details on device fabrication, characterization and measurement configuration are provided in the Methods, Supplementary Note 1 and 2 and figs. S1 and S2. All measurements were performed at $T = 10$ mK, unless otherwise specified.

Fig. 1C displays the phase diagram under a large negative $D$ that polarizes conduction band away from the moiré superlattice interface, which exhibits a quantized Hall resistance $R_{yx} \approx h/e^2$ accompanied by a vanishing longitudinal resistance $R_{xx}$ (fig. S2D) at a moiré band filling factor $v = 1$ and $B_\perp = 50$ mT. Consistent with this observation, Hartree-Fock calculations (Supplementary Note 8.1) obtain a spin-valley polarized ground state hosting an isolated occupied conduction band with $|C| = 1$ (Fig. 1B), similar to previous theoretical results in other rhombohedral multilayer graphene (*60-65*). The Landau fan diagram at $D = -0.69$ V/nm (Fig. 1D) exhibits the QAH state with switchable $C = \pm 1$. The $R_{yx}$ vs $B_\perp$ loop at $v = 1$ and $D = -0.69$ V/nm (Fig. 1F) reveals a QAH state with counter-clockwise chirality and coercive field $B_C \sim 30$ mT, and the $v$ sweep in Fig. 1H shows the quantization of $R_{yx}$ and vanishing $R_{xx}$ around $v = 1$. Surprisingly, when the displacement field is tuned to $D = -0.64$ V/nm (Fig. 1E), the $C = +1$ QAH state recedes, leaving only the $C = -1$ QAH state near $v = 1$. Meanwhile, the $R_{yx}$ vs $B_\perp$ loop at $v = 1$ (Fig. 1G) reveals a QAH state with clockwise chirality and coercive field $B_C \sim 170$ mT. The Landau fan diagrams at other displacement fields (fig. S3) reveal the systematic disappearance of the $C = +1$ QAH state at positive $B_\perp$ on the right-hand side of the phase diagram, leaving only the $C = -1$ QAH state on the left-hand side. The critical positive $B_\perp$ ($B_{T\perp}$) for the transition between the two sets of QAH state can be extracted from the black dashed lines in the Landau fan diagrams (fig. S3), which gradually decreases with decreasing $D$ (Fig. 1I).

**Chern number reversal of the QAH state induced by $B_\parallel$**

When $B_\parallel$ is applied to the 8L-RG at $D = -0.69$ V/nm and $v = 0.98$, the QAH effect evolves in a highly systematic yet non-monotonic manner. As $B_\parallel$ rises from 0 to 0.1 T (blue curves in Fig. 2A), the out-of-plane hysteresis loops retain the same chirality while $B_C$ increases from $\sim 30$ mT to $\sim 170$ mT. With further increase of $B_\parallel$ between 0.2 T and 1 T (red curves), however, the QAH chirality reverses while $B_C$ decreases from $\sim 470$ mT to $\sim 130$ mT. Throughout the whole range of $B_\parallel$, the QAH state remains well quantized. The evolution of Landau fan diagrams with $B_\parallel$ reveals a progressive suppression of the $C = +1$ QAH phase with increasing $B_\parallel$ and a transition to the $C = -1$ state. The $B_{T\perp}$ (Fig. 2E) extracted for each $B_\parallel$ is indicated by black dashed lines in Fig. 2, B and C and fig. S4, which exhibit a similar trend to that modulated by $D$ in Fig. 1I. To elucidate the role of $B_\parallel$ in regulating the QAH state, we investigated the angular dependence of magnetic hysteresis loops by rotating a $B_\parallel = 0.6$ T with angle $\theta_B = 0°$-$330°$ in the plane, which almost all collapse to the same behavior (Fig. 2F). We extracted the magnetic hysteresis window $\Delta B_C$ between the positive and negative coercive fields, as illustrated in Fig. 2G, which remains nearly isotropic with respect to $\theta_B$. The Landau fan diagrams at different $B_\parallel$ also display isotropic characteristics (figs. S5 and S6), as well as the results at another $D$ (fig. S7). This observation suggests that the variation of QAH state under $B_\parallel$ is not dominated by in-plane orbital magnetism, which is expected to produce an anisotropic response (*66-*

67). To rule out the possible artifact associated with the small $B_\perp$ component caused by unavoidable misalignment of $B_\parallel$ relative to the in-plane direction, we developed a strict procedure described in Supplementary Note 6 and fig. S8.

We also observe the quantized hysteresis loop of the QAH state modulated purely by $B_\parallel$, as shown in Fig. 2H. The hysteresis loop versus $B_\parallel$ exhibits a single stable quantized chiral state at zero field, as well as two independent and symmetric hysteresis windows on either side. These loops can also be tuned by $B_\perp$, which causes both the chirality reversal and the non-monotonic modulation of hysteresis windows. When the negative $B_\perp$ increases from -0.4 T to -0.2 T, the hysteresis windows on both sides of zero field gradually expand. In contrast, a positive $B_\perp \sim 0.2$ T reverses the chirality of the QAH ground state. As the positive $B_\perp$ further increases to 0.4 T, the hysteresis windows on both sides progressively shrink. In Fig. 2, I to K and fig. S9, we map $R_{yx}$ as a function of $v$ and $B_\parallel$ at several fixed $B_\perp$, which directly visualizes the evolution of QAH states under the combined influence of $B_\parallel$ and $B_\perp$.

**Three moiré-proximal superconducting phases**

The QAH state described above was observed in the electron-doped regime with carriers away from the moiré superlattice. Strikingly, when $v$ is tuned toward the hole-doped regime with carriers near the moiré superlattice, two SC phases emerge around $v = -1$ and $v = -4$, as revealed by the two white crescents denoted as SC1 and SC2 in Fig. 3A. The finely swept $v$–$D$ phase diagram is presented in fig. S2, E and F. The inset of Fig. 3D shows that the resistance drops to very close to zero in the $v$ range within the SC phase, and the line cut along the red dashed line in Fig. 3A shows that the QAH and SC states are linked by metallic phases in between. More surprisingly, the application of $B_\parallel$ gives rise to a new SC state near $v = -2$, as revealed by the third white crescent denoted as SC3 in Fig. 3B. At $D = -0.19$ V/nm, line cuts across the SC3 regime in Fig. 3E show finite resistance without $B_\parallel$, whereas a zero-resistance SC state emerges under $B_\parallel = 0.5$ T. Upon increasing the field to $B_\parallel = 0.9$ T (Fig. 3C), SC1 and SC3 merge into a connected sliver in the $v$-$D$ phase diagram, whereas SC2 remains separated with a reduced phase space.

Temperature-dependent transport measurements (Fig. 3F) show that SC1, SC2, and SC3 exhibit SC transitions at $T_C \sim 120$ mK, 100 mK and 88 mK (defined as the temperature at which $R_{xx}$ reaches 50% of the normal-state value). All three SC phases are rapidly suppressed by a perpendicular magnetic field, as revealed by the $dV/dI$ vs $B_\perp$ maps in Fig. 3, G to I and spectral evolution in Fig. 3, J to L. The upper critical fields $B_{C2,\perp}$ determined from the $dV/dI$ maps are $\sim 8$ mT and 6 mT for SC1 and SC2 at $B_\parallel = 0$, and 20 mT for SC3 at $B_\parallel = 0.5$ T (Fig. 3, G to I). The coherence length $\xi$ can be estimated from the relation $B_{C2,\perp} = \Phi_0/2\pi\xi^2$, where $\Phi_0$ is the superconducting flux quantum, yielding $\xi \sim$ 203 nm, 234 nm, and 128 nm for the three cases, respectively.

**Distinctively different responses of the SC states to $B_\parallel$**

The three SC phases not only reside in different regions of the phase diagram, but also exhibit distinctively different responses to $B_\parallel$, as depicted in Fig. 4, A to C. With increasing $B_\parallel$ at a fixed $D$, the SC1 region in the $v$–$B_\parallel$ phase diagram gradually expands to a slightly larger $v$ range. In contrast, the SC2 region shrinks with increasing $B_\parallel$ and completely vanishes for $B_\parallel \sim 0.9$ T. Notably, SC3 emerges only at $B_\parallel > 0.2$ T, and its phase space expands significantly with increasing $B_\parallel$ up to $\sim 0.9$ T. We then track the d$V$/d$I$ vs $I_{dc}$ curves as a function of $B_\parallel$ along the red dashed lines in Fig. 4, A to C. The traces for SC1 are insensitive to $B_\parallel$, the SC2 state is gradually suppressed by $B_\parallel$, and the SC3 state emerges under $B_\parallel$ and becomes more robust with increasing $B_\parallel$ (Fig. 4, D to F). For a weak-coupling spin-singlet BCS superconductor, the Pauli paramagnetic limit for Zeeman-effect-induced pair breaking is given by $B_P = 1.86$ T/K $\times$ $T_C$. Based on the $T_C$ values estimated above, the Pauli limits for the three SC phases are $\sim 0.2$ T. However, all three SC states survive in a $B_\parallel$ strength significantly exceeding the Pauli limit, strongly indicating unconventional superconductivity. Meanwhile, the SOC strength is estimated to be $\sim 60$ μeV (Supplementary Note 8.2 and fig. S10), leading to the protection against $B_\parallel \sim 0.5$ T, which is substantially smaller than the critical $B_\parallel$ of the three SC states.

We have also measured the evolution of d$V$/d$I$ vs $I_{dc}$ curves in SC3 with $v$ (Fig. 4G), and its angular dependence under $B_\parallel$ at a fixed $v = -2.5$ (Fig. 4H). The data reveal that SC3 exhibits an isotropic response to the in-plane field (Fig. 4I), which effectively rules out an in-plane orbital origin, which is expected to exhibit a $C_3$-symmetric threefold anisotropy (*68-70*). This behavior instead suggests that $B_\parallel$ couples directly to spin, while any additional influence arises indirectly through SOC. We can therefore exclude the Fulde–Ferrell–Larkin–Ovchinnikov (FFLO) states as in NbSe$_2$, where the in-plane orbital effects are crucial and typically exhibit sixfold symmetry (*71-73*).

**Discussions and Conclusions**

Below we will give a coherent explanation for the QAH Chern number reversal and three unconventional SC phases in the presence of $B_\parallel$ in the 8L-RG moiré superlattice.

For the field control of QAH state, we first note that it is spin-valley-polarized and the Chern number reversal can be understood as the competition between different spin-valley polarized states. Because $B_\parallel$ mainly acts on the spin magnetic moment, SOC is essential for coupling $B_\parallel$ to the valley degree of freedom (*74*). On the other hand, $B_\perp$ influences the valley polarization $\tau_z$ both through the orbital term $-B_\perp M_z \tau_z$ and the SOC term $-\lambda_I \tau_z s_z /2$ since $B_\perp$ modifies $s_z$. While the orbital contribution grows linearly with $B_\perp$, the SOC term quickly saturates as $|s_z|$ rapidly approaches unity in a finite $B_\perp$. Therefore, SOC dominates at low field, and the orbital term dominates at high field. If they favor opposite valley-polarization, $B_\perp$ can reverse the Chern number at a characteristic $B_{T\perp} = \lambda_I/2M_z$. Consistent with experimental observation, $B_{T\perp}$ will be reduced by $B_\parallel$, which cants the spins into the plane and suppresses $s_z$, thereby weakening the effective SOC. The displacement field primarily tunes $M_z$ through the electron wavefunction, which also affects $B_{T\perp}$ as revealed by Fig. 1I. By solving the effective Hamiltonian including the effects of $B_\perp$, $B_\parallel$ and SOC, we obtain a simple relation of the critical $B_\perp$ and $B_\parallel$ that matches well with the experimental phase boundary at various fillings (Supplementary

Note 8.2 and fig. S10A). The extracted SOC strength is around 60 μeV, in good agreement with previous experimental values of 40-120 μeV in multilayer graphene (*42,75-78*), and is nearly filling-independent as expected (fig. S10B).

The hysteresis tuned by $B_∥$ and $B_⊥$ can also be understood from the competition between $M_z$ and SOC. For $B_⊥ > 0$, SOC favors the $C = +1$ state over the $C = -1$ state as discussed above, and increasing $B_∥$ suppresses the effective SOC. As a result, the $C = -1$ region always expands with increasing $B_∥$ for positive $B_⊥$, consistent with Fig. 2A. In contrast, increasing $B_⊥$ can either stabilize or destabilize the $C = +1$ state relative to the $C = -1$ state, depending on whether the orbital magnetic contribution or the increase of spin polarization is dominant, also consistent with Fig. 2H. To capture these behaviors, we formulate a Landau free energy and simulate the evolution of hysteresis with $B_∥$ and $B_⊥$, as detailed in Supplementary Note 8.3 and figs. S11 and S12. The calculation reproduces the key features of Fig. 2, A and H.

Next, we discuss the possible pairing mechanisms of the three SC states. The SC2 state survives the Pauli limit, but exhibits rather conventional suppression by $B_∥$. It is most likely because SC2 emerges at $v < -4$, far away from the flat band thus the strong-correlation effect diminishes. The SC2 pairing is likely dominated by spin-singlet (↑↓−↓↑) and exhibits Ising-like superconductivity. The SC1 and SC3 phases, in contrast, emerge near 1/4 filling and half-filling of the moiré flat bands, where strong-correlation effect dominates. Under an $B_∥ = 0.9$ T, SC1 and SC3 are connected in the $v$–$D$ parameter space, suggesting that they share similar origin. The SOC in this system, which tends to mix the spin singlet and triplet pairing, is much weaker than the Hund's rule coupling (*54,79-81*) and electron-phonon-mediated interaction (*82-83*), which generally split the singlet and triplet channels and are typically of order meV. Therefore, the singlet-triplet mixing is expected to remain weak unless the two channels are nearly degenerate accidentally. Under this assumption, the singlet and triplet channels can be discussed separately. Since spin-singlet pairing is generally disfavored by a magnetic field, the emergence of SC3 and the slight expansion of SC1 under finite $B_∥$ can be naturally interpreted as evidence that the field stabilizes spin-triplet pairing. The triplet pairing (if it exists) may be either $S_z = 0$ (↑↓+↓↑) or equal-spin pairing (↑↑, ↓↓) in $z$ direction at zero $B_∥$. With increasing $B_∥$, the spin configuration is gradually canted towards the equal-spin triplet state polarized along the $x$ direction, thereby gaining spin Zeeman energy. The fact that SC1 exists at zero field while SC3 only emerges at finite $B_∥$ suggests that the zero-field region that develops into SC3 lies close to a pairing instability, but the associated superconducting state remains slightly unfavorable in energy there. An in-plane magnetic field can lower the energy of the in-plane spin configuration, thereby triggering this pairing instability.

The coexistence of Chern number reversible QAH state and three unconvetional SC phases highlights the crucial role of spin degree of freedom in rhombohedral graphene moiré superlattices. We also identify the in-plane magnetic field as a powerful in-situ control knob that can precisely manipulate each quantum phase and the interaction between them for technological applications. Together, these results establish a versatile platform for exploring exotic quantum phenomena at the intersection of topology and superconductivity, such as the Majorana zero mode and topological quantum computing.

**Data availability**

The data that support the findings of this study are available from the corresponding authors on a request.

**Acknowledgments**

We thank Jianpeng Liu and Min Li for valuable discussions. We thank Xian Wang and Xiaoqing Xi for technical support with the scanning near-field optical microscopy measurements. We thank the Synergetic Extreme Condition User Facility (SECUF) for support. Yayu Wang acknowledges supports from the Basic Science Center Project of NSF of China (Grant No. 52388201), the Innovation program for Quantum Science and Technology (Grant No. 2021ZD0302502) and the New Cornerstone Science Foundation through the New Cornerstone Investigator Program. K.W. and T.T. acknowledge support from the CREST (JPMJCR24A5), JST and World Premier International Research Center Initiative (WPI), MEXT, Japan. Zhida Song acknowledges supports from NSF of China (Grant No. 12274005), National Key Research and Development Program of China (No. 2021YFA1401900), and Innovation program Quantum Science and Technology (No. 2021ZD0302403). Jinsong Zhang acknowledges supports from the National Key R&D Program of China (Grant No. 2024YFA1409100), the NSF of China (Grant No. 12274252, 12350404). Jie Shen acknowledge supports from the Beijing NSF (Grant No. JQ23022) and the National Key Research and Development Program of China (Grant Nos. 2023YFA1607400 and 2024YFA1600042).

**Author contributions**

Y.W. supervised the research project. X.Z. and Y.W. designed the experiments and analyzed the data; X.Z. and J.G. fabricated the devices with assistance from K.C., C.G., and Z.X.; X.Z. and H.L. performed the transport measurements; Z.S. proposed the SOC mechanism theory; G.Z. performed the numerical calculations; A.W., J.S., and J.Z. provided measurement support; K.W. and T.T. provided hBN crystals; X.Z., G.Z. Z.S. and Y.W. wrote the manuscript with the input from all authors.

**Competing interests**

The authors declare no competing interests.

**Figure Captions**

**Fig. 1. QAH phase diagram of 8L-RG moiré superlattice.** (**A**) Schematic illustration of an hBN-encapsulated, dual-gated 8-layer rhombohedral graphene moiré superlattice device with the top-gate voltage ($V_{tg}$) and bottom-gate voltage ($V_{bg}$) applied through graphite gates. The twist angle between the 8L-RG and the top hBN is 0.55°. (**B**) Hartree-Fock band structure of the spin-valley-polarized state at $v = 1$ and $D = -30$ meV. The Fermi level is set at zero energy. The occupied conduction band above the charge neutrality point has Chern number $|C| = 1$ and is highlighted in red. (Inset) Berry curvature $\Omega(k)$ of the corresponding band. Here $\Omega(k)$ is normalized such that a band with $C = 1$ has an average Berry curvature of $2\pi$. (**C**) Color plot of the Hall resistance $R_{yx}$ as functions of $D$ and $v$ at $B_\perp = 50$ mT. (**D** and **E**) Landau fan diagrams of $R_{yx}$ measured at fixed $D = -0.69$ V/nm (D) and $D = -0.64$ V/nm (E), respectively. White dashed lines show the expected evolution of Chern number $C = \pm 1$ QAH states based on the Streda's formula $\partial n / \partial B_\perp = C\, e/h$. The black dashed line marks the critical positive $B_\perp$ ($B_{T\perp}$) for the phase transition from the $C = +1$ to the $C = -1$ QAH state. (**F** and **G**) Out-of-plane magnetic hysteresis loops of $R_{xx}$ and $R_{yx}$ measured at $v = 0.98$ with $D = -0.69$ V/nm (F) and $D = -0.64$ V/nm (G). The QAH states exhibit opposite chiralities with $B_C \sim 30$ mT (F) and $\sim 170$ mT (G). (**H**) The $R_{xx}$ and $R_{yx}$ values versus $v$ measured at 0 T and $D = -0.69$ V/nm, showing quantized $R_{yx} = h/e^2$ and vanishing $R_{xx}$. (**I**) $B_{T\perp}$ for the transition from the $C = +1$ to the $C = -1$ QAH state, extracted from the Landau fan diagrams of $R_{yx}$ at different $D$ and $B_\parallel = 0$ T. The corresponding values are indicated by the black dashed lines in Fig. 1,D and E, and fig. S3.

**Fig. 2. Chern number reversal induced by $B_\parallel$.** (**A**) Out-of-plane magnetic hysteresis loops measured under different $B_\parallel$ at $D = -0.69$ V/nm and $v = 0.98$. Solid (dashed) lines correspond to sweeping $B_\perp$ from positive (negative) values to negative (positive) values. For small $B_\parallel = 0$ T, 25 mT, 50 mT, and 0.1 T, the corresponding $B_C$ are $\sim 0.03$ mT, $\sim 0.09$ T, $\sim 0.12$ T, and $\sim 0.17$ T, respectively. As $B_\parallel$ is further increased, the chirality of the QAH state reverses. For larger $B_\parallel = 0.2$ T, 0.4 T, 0.6 T, 0.8 T, and 1 T, the measured $B_C$ are $\sim 0.47$ T, $\sim 0.36$ T, $\sim 0.19$ T, $\sim 0.16$ T, and $\sim 0.13$ T, respectively. (**B** to **D**) Landau fan diagrams of $R_{yx}$ at $D = -0.69$ V/nm measured under $B_\parallel = 0.2$ T (B), 0.4 T (C) and 0.6 T (D), respectively. (**E**) The $B_{T\perp}$ for the transition from the $C = +1$ to the $C = -1$ QAH state, extracted from the Landau fan diagrams of $R_{yx}$ at different $B_\parallel$ and $D = -0.69$ V/nm. The corresponding values are indicated by the black dashed lines in Fig. 2, B and C, and fig. S4. (**F**) Out-of-plane magnetic hysteresis loops measured at $D = -0.69$ V/nm, $v = 0.98$ and $B_\parallel = 0.6$ T, under different in-plane rotation angles $\theta_B = 0°$-330°. (**G**) The out-of-plane magnetic hysteresis window $\Delta B_C = B_C + |-B_C|$ of the QAH state was extracted from the hysteresis loops in Fig. 2F, showing its dependence on $\theta_B$. (**H**) In-plane magnetic hysteresis loops at $B_\perp = -0.4$ T, -0.2 T, 0.2 T, 0.3 T and 0.4 T, measured at $D = -0.69$ V/nm and $v = 1.01$. (**I** to **K**) Maps of $R_{yx}$ as functions of $v$ and $B_\parallel$ at $D = -0.69$ V/nm, measured under $B_\perp = 0.2$ T (I), 0.4 T (J) and -0.4 T (K).

**Fig. 3. Superconductivity near the moiré superlattice.** (**A** to **C**) Color maps of $R_{xx}$ versus $D$ and $v$ at $B_{\parallel} = 0$ T (A), 0.5 T (B) and 0.9 T (C), respectively. The white crescents represent zero-resistance superconducting states. (**D**) Line cuts of $R_{xx}$ (blue line) and $R_{yx}$ (red line) as a function of $v$ along the red dashed line in Fig. 3A. (Inset) $R_{xx}$ plotted against $v$, with the black dashed line indicating zero resistance. (**E**) $R_{xx}$ versus $v$ measured at $B_{\parallel} = 0$ (blue line) and 0.5 T (red line) with $D = -0.19$ V/nm, revealing that the SC3 phase is induced by in-plane magnetic fields. (**F**) Temperature-dependent $R_{xx}$ traces of SC1, SC2, and SC3, at $v = -0.8$ and $D = -0.31$ V/nm (black lines); $v = -4.7$ and $D = -0.07$ V/nm (red lines); $v = -2.5$ and $D = -0.19$ V/nm (blue lines). The corresponding $T_C$ are 120 mK, 100 mK and 88 mK, respectively. (**G** to **I**) Color maps of $dV/dI$ for SC1, SC2, and SC3 as a function of DC bias current $I_{dc}$ at $v = -0.8$, $D = -0.31$ V/nm and $B_{\perp} = 0$ T (G); $v = -4.7$, $D = -0.07$ V/nm and $B_{\perp} = 0$ T (H); and $v = -2.5$, $D = -0.19$ V/nm and $B_{\perp} = 0.5$ T (I). Perpendicular upper critical fields for SC1, SC2, and SC3 are extracted to be $B_{C2,\perp} = 8$ mT, 6 mT, and 20 mT, respectively. (**J** to **L**) Line cuts extracted from the color maps (G to I) at representative $B_{\perp}$ values for SC1 (J), SC2 (K), and SC3 (L). All curves are vertically shifted for better visualization.

**Fig. 4. Responses of the three SC states to $B_{\parallel}$.** (**A** to **C**) Maps of $R_{xx}$ for SC1 (A), SC2 (B) and SC3 (C) as functions of $v$ and $B_{\parallel}$, with $D$ fixed at -0.31 V/nm, -0.07 V/nm and -0.19 V/nm, respectively. With increasing $B_{\parallel}$, the SC1 phase is slightly expanded, SC2 is gradually suppressed, and SC3 is induced and further enhanced. (**D** to **F**) Color maps of $dV/dI$ versus $I_{dc}$ and $B_{\parallel}$ for SC1 (D), SC2 (E) and SC3 (F), measured along the red dashed lines in Fig. 4, A to C, respectively. The three SC phases also exhibit distinctively different responses to $B_{\parallel}$. (**G**) Color map of $dV/dI$ in SC3 versus $v$ and $I_{dc}$, at $D = -0.19$ V/nm and $B_{\parallel} = 0.5$ T, showing that superconductivity is strongest at $v = -2.5$. (**H**) Line cuts of $dV/dI$ vs $I_{dc}$ for SC3 at $D = -0.19$ V/nm, $v = -2.5$ and $B_{\parallel} = 0.5$ T, under different in-plane rotation angles $\theta_B = 0°$-$180°$. All the curves are almost identical, revealing isotropic response to $B_{\parallel}$. (**I**) The superconducting critical current difference $\Delta I_C = I_C + |-I_C|$ is extracted from Fig. 4H and plotted as a function of $\theta_B$. Data from 0° to 180° are measured and data from 180° to 360° are obtained by symmetric extension.

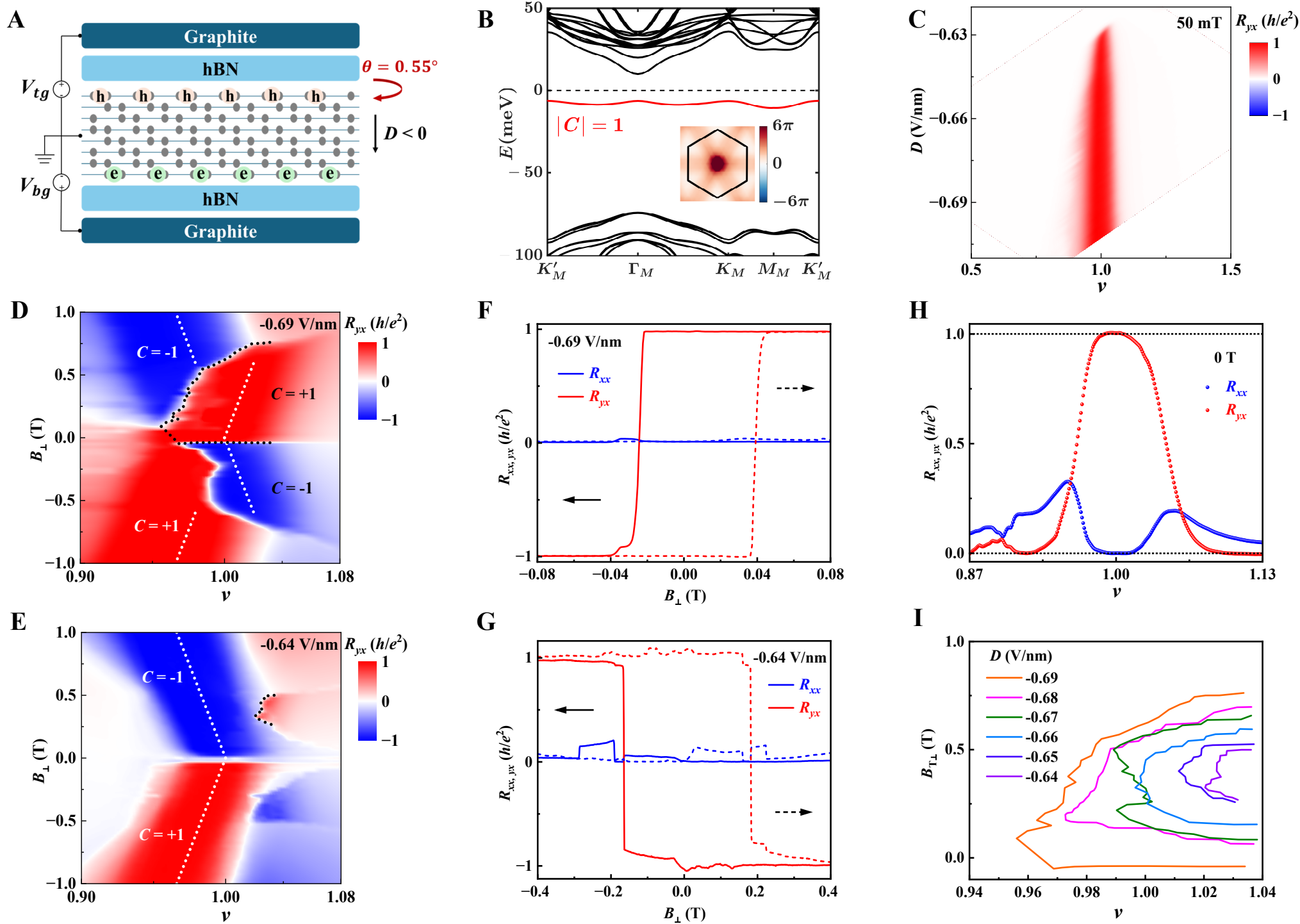

**Figure 1**

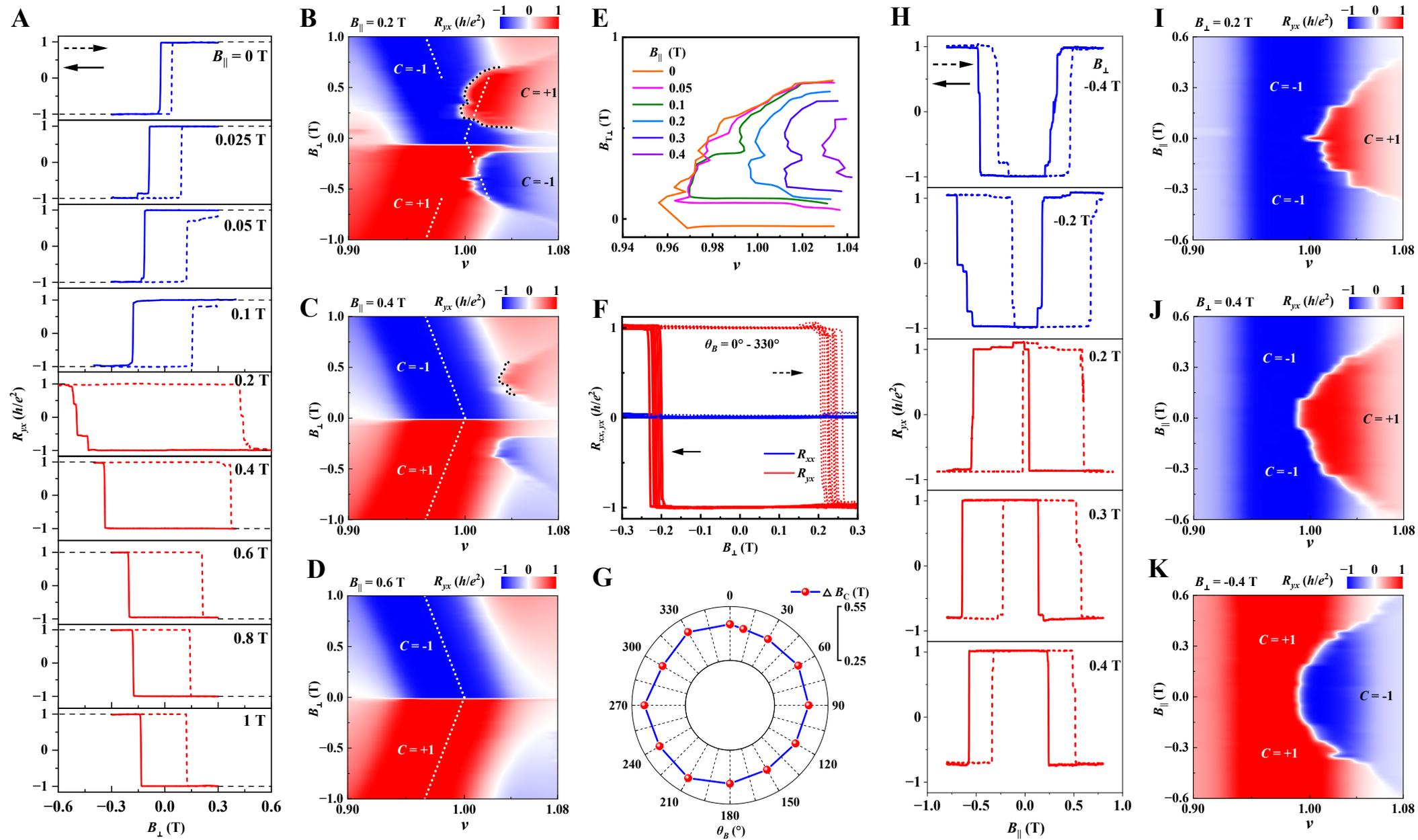

Figure 2

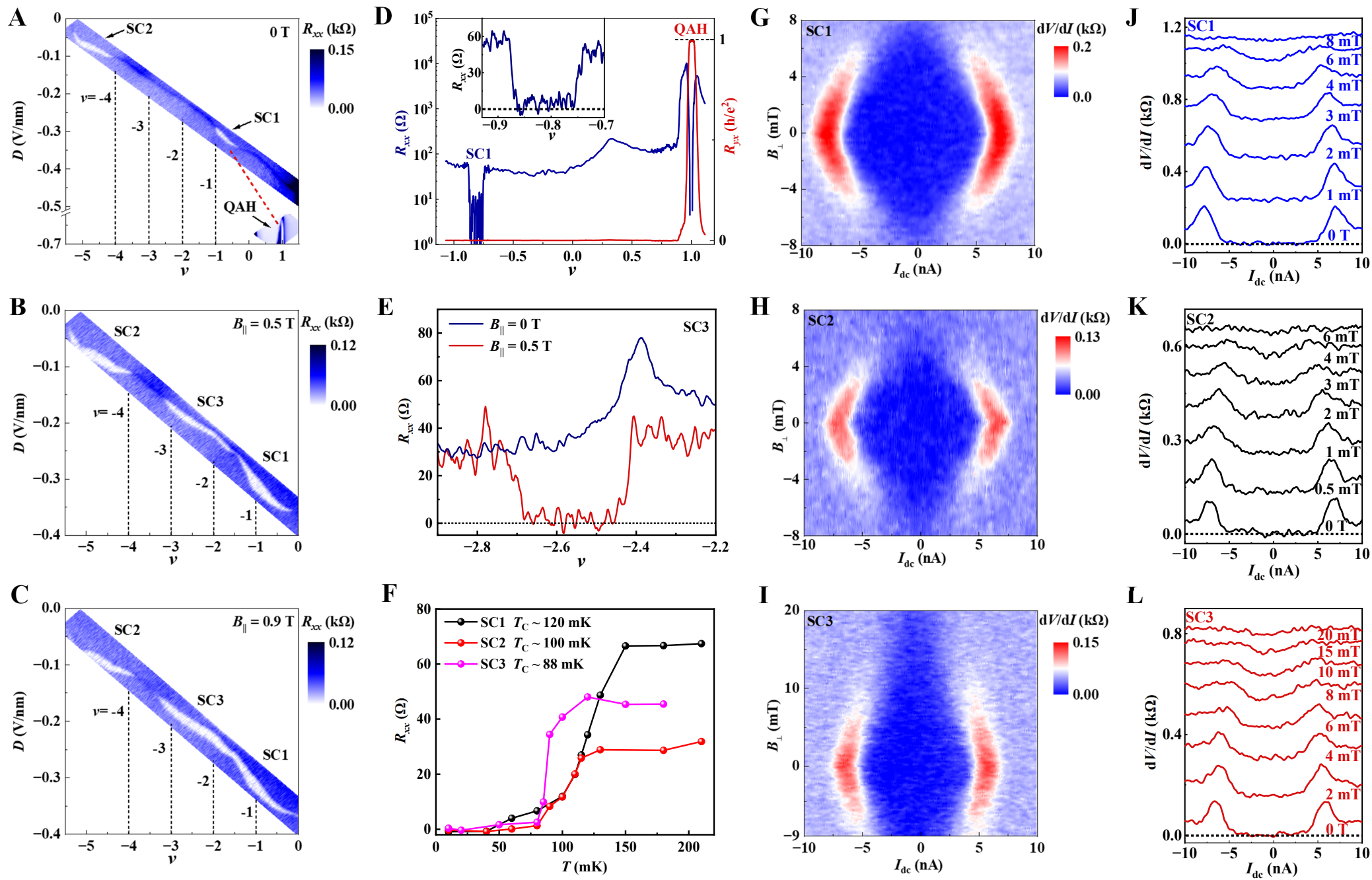

Figure 3

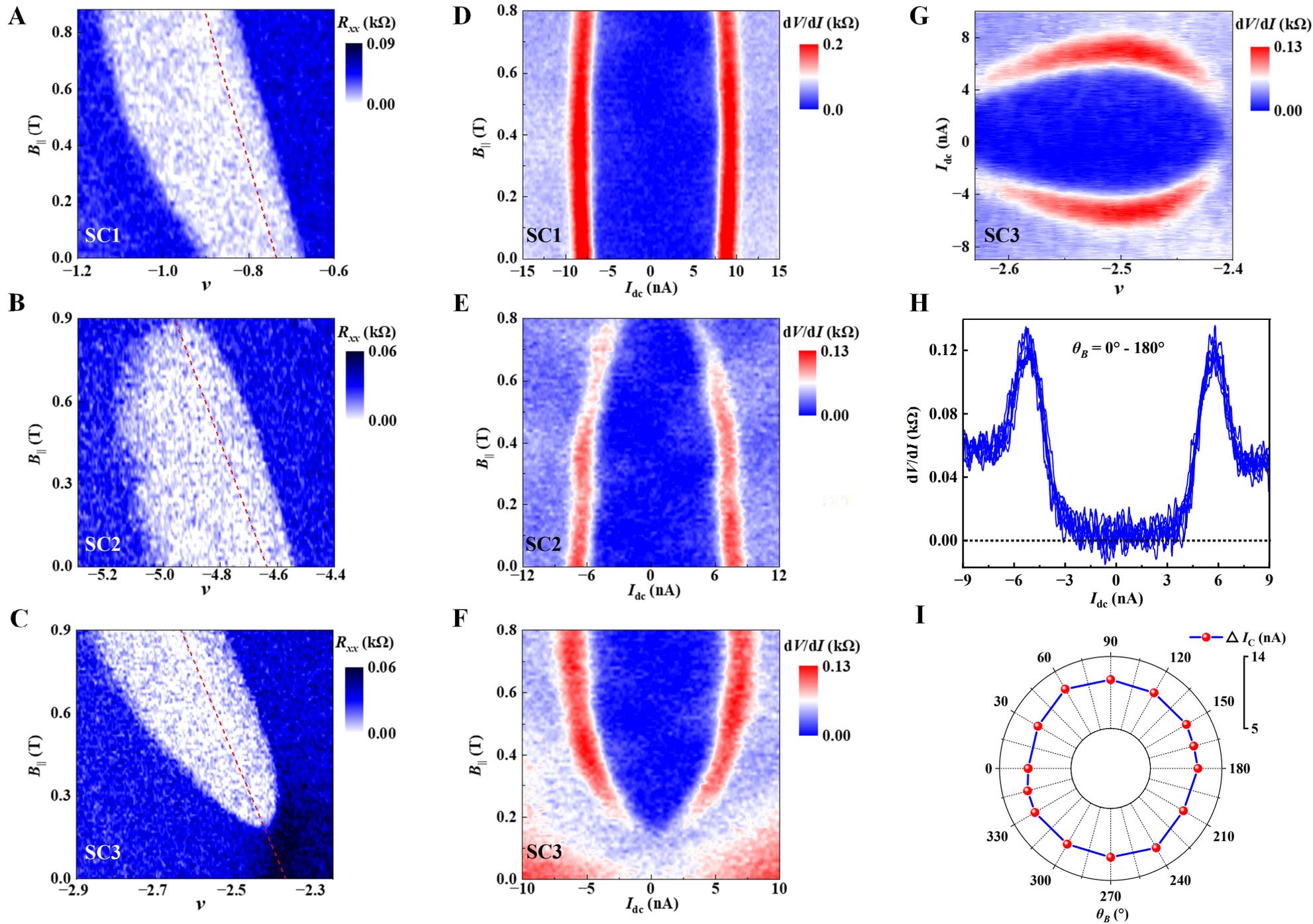

Figure 4

# Supplementary Materials for

# Chern number reversal and emergent superconductivity in rhombohedral graphene induced by in-plane magnetic fields

Xiaozhou Zan[1,10]*, Hangzhe Li[1,10], Jiawei Guo[1,10], Gengdong Zhou[2,10], Kangyao Chen[1], Cihan Gao[1], Zijun Xu[1], Kenji Watanabe[3], Takashi Taniguchi[4], Anqi Wang[5], Jie Shen[5,6], Jinsong Zhang[1,7,9], Zhida Song[2,7,8]* and Yayu Wang[1,7,9]*

[1]State Key Laboratory of Low Dimensional Quantum Physics, Department of Physics, Tsinghua University, Beijing 100084, China.

[2]International Center for Quantum Materials, School of Physics, Peking University, Beijing 100871, China.

[3]Research Center for Electronic and Optical Materials, National Institute for Materials Science, 1-1 Namiki, Tsukuba 305-0044, Japan.

[4]Research Center for Materials Nanoarchitectonics, National Institute for Materials Science, 1-1 Namiki, Tsukuba 305-0044, Japan.

[5]Beijing National Laboratory for Condensed Matter Physics, Institute of Physics, Chinese Academy of Sciences, Beijing 100190, China.

[6]School of Physical Sciences, University of Chinese Academy of Sciences, Beijing 100190, China.

[7]Hefei National Laboratory, Hefei 230088, China.

[8]Beijing Key Laboratory of Quantum Devices, Peking University, Beijing 100871, China.

[9]New Cornerstone Science Laboratory, Frontier Science Center for Quantum Information, Beijing 100084, China.

[10]These authors contributed equally: Xiaozhou Zan, Hangzhe Li, Jiawei Guo, Gengdong Zhou

*Corresponding authors: *zanxz@tsinghua.edu.cn; songzd@pku.edu.cn; yayuwang@tsinghua.edu.cn*

**The PDF file includes:**

    Materials and Methods
    Supplementary Note 1 to 9
    Figs. S1 to S13
    References

## Materials and Methods

### Devices fabrication

A combination of scanning near-field optical microscopy (SNOM), Raman spectroscopy, and a continuous-wave laser enabled the identification and segmentation of rhombohedral stacking order. Following this identification, high-quality, hexagonal boron nitride (hBN)-encapsulated, dual-graphite-gated devices were fabricated via a dry transfer technique in which a poly(Bisphenol A carbonate) (PC) film on a polydimethylsiloxane (PDMS) support was used to sequentially pick up the two-dimensional materials. Finally, the Hall bar devices were fabricated using standard electron-beam lithography (EBL), electron-beam metal evaporation, and reactive ion etching (RIE) with a $CHF_3$ and $O_2$ gas mixture. Details of the sample fabrication process are provided in Supplementary Note 1 and fig. S1.

### Transport measurements

Electrical transport properties were measured in a dilution refrigerator equipped with vector magnets providing a vertical field up to 6 T and in-plane fields of 1 T along $x$ and 1 T along $y$, with a base temperature of 10 mK. Using d$V$/d$I$ maps of the spin-polarized SC3 state versus $B_\perp$, we quantitatively determined the tilt components of $B_\parallel$ along the $x$ and $y$ axes (both <1°, see Supplementary Note 6 and fig. S8), and performed precise vector magnet calibration to fully eliminate any residual out-of-plane component of the in-plane field. The $V_{tg}$ and $V_{bg}$ gate voltages were applied through two Keithley 2400 source meters. The excitation current was provided by a Keithley 6221 current source with an excitation of 1 nA at a frequency of 11.777 Hz. Longitudinal resistance and Hall resistance were measured using standard lock-in technique. Note that the quantized $R_{yx}$ and zero $R_{xx}$ values in this work are presented without any symmetrization or antisymmetrization processing.

The carrier density $n$ and displacement field $D$ were tuned by sweeping the $V_{tg}$ and $V_{bg}$, following the relations: $n = (C_b V_{bg} + C_t V_{tg})/e$ and $D = (C_b V_{bg} - C_t V_{tg})/2\varepsilon_0$, where $e$ is the elementary charge, $\varepsilon_0$ denotes the vacuum permittivity, $C_t$ and $C_b$ are top gate and bottom-gate capacitance per area calculated from the Landau fan diagram. The twist angle ($\theta$) between graphene and hBN can be estimated from the carrier density at full filling $n_s$ of a moiré band. The relationship is described by $n_s \approx 8\theta^2/\sqrt{3}a^2$, where $a = 0.246$ nm is the lattice constant of graphene. And the corresponding moiré period $\lambda$ is described by $\lambda = \frac{(1+\delta)a}{\sqrt{2(1+\delta)[1-\cos(\theta)]+\delta^2}}$, where the $\delta \approx 1.7\%$ represents lattice mismatch between hBN and graphene (*1-3*).

## Supplementary Note 1. 8L-RG moiré superlattice device fabrication

The graphene and hBN flakes were mechanically exfoliated on 285 nm-$SiO_2$/Si substrates. Regions of octalayer graphene with rhombohedral (ABC) stacking order were identified by SNOM with an excitation frequency of 950 cm$^{-1}$ in fig. S1A. The stacking order was further confirmed by Raman spectroscopy excited by 532 nm (2.33 eV) laser, and the targeted ABC domain was subsequently defined by continuous-wave laser cutting in fig. S1B. The red circle in the optical microscope image indicates the position of the Raman laser measurement. The 2D mode of ABC graphene including a sharp peak on the left and a flat shoulder on the right, consistent with previous observations (*4*). Fig. S1C shows the second SNOM characterization to verify that the stacking order in the isolated region is solely ABC stacking order after cutting.

We fabricated the van der Waals heterostructure using a dry transfer method by sequentially picking up the following layers: top hBN, the target octalayer ABC-stacked graphene, bottom hBN, and a bottom graphite gate, and then transferring the complete stack onto a SiO$_2$/Si substrate. A twist angle was introduced between the top hBN and the octalayer graphene during the dry transfer process. Fig. S1D shows a schematic diagram and an optical microscope image of the heterostructure. To confirm stacking order preservation, Raman spectroscopy was performed on the transferred heterostructures and confirmed that the rhombohedral stacking order remains unchanged in fig. S1E. Finally, a top graphite gate was transferred onto the heterostructure, which was then patterned into a standard Hall bar geometry using micro/nanofabrication techniques. Fig. S1F presents optical micrographs of 8L-RG moiré devices D1 and D2.

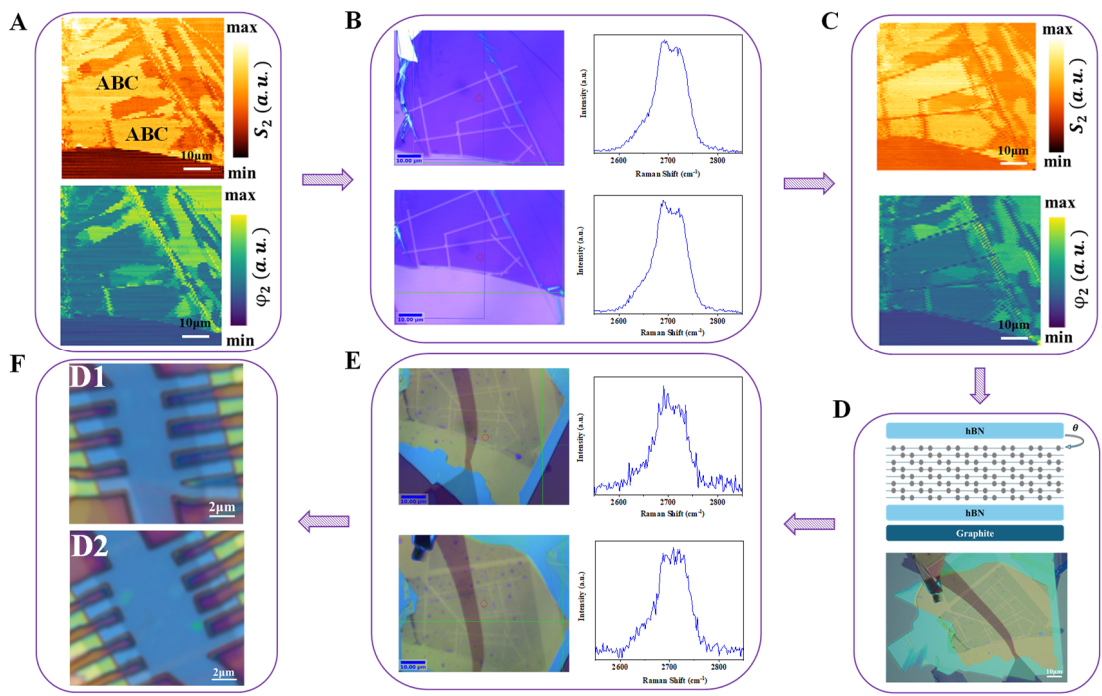

**Fig. S1. Device fabrication procedure.** (**A**) The near-field amplitude ($S_2$) and phase ($\varphi_2$) images with an excitation frequency of 950 cm$^{-1}$ of octalayer graphene. The rhombohedral (ABC) stacking order is identified and annotated in the figure. (**B**) The optical image and Raman spectrum excited by 532 nm (2.33 eV) laser of graphene after laser cutting. The red circle in the optical microscope image indicates the point of the Raman laser measurement. (**C**) The near-field amplitude ($S_2$) and phase ($\varphi_2$) images

with an excitation frequency of 950 cm$^{-1}$ after laser cutting demonstrate that the processed area is composed solely of rhombohedral stacking order. **(D)** Schematic and optical micrograph of the van der Waals heterostructure. **(E)** Optical micrograph and Raman spectrum excited by 532 nm (2.33 eV) laser of the van der Waals heterostructure. The red circle in the optical microscope image indicates the point of the Raman laser measurement. **(F)** optical micrographs of devices D1 and D2.

## Supplementary Note 2. The $v$–$D$ phase diagram of 8L-RG moiré superlattice

The measurement configuration for $R_{xx}$ and $R_{yx}$ under a vector magnetic field is presented in fig. S2A. $B_∥$ and $B_⊥$ denote magnetic fields parallel and perpendicular to the two-dimensional plane, respectively. And the $B_∥$ is composed of in-plane components $B_x$ and $B_y$, with a rotation angle defined as $θ_B$. The $R_{xx}$ features at $n ≈ 2.85×10^{12}$ cm$^{-2}$ and $n ≈ 1.43×10^{12}$ cm$^{-2}$ in fig. S2B are identified as full filling ($v = 4$) and half filling ($v = 2$) of the moiré unit cell. From these fillings, the twist angle between the 8L-RG and the top-layer hBN is determined as $θ ≈ 0.55°$, yielding a moiré period of $λ ≈ 12.7$ nm (*1-3*). SC1 and SC2 are only observed in regions adjacent to the moiré superlattice, specifically within the hole-doped area and distributed along the same resistance ridge in the phase diagram. No SC features can be identified in other regions (fig. S2C). The enlarged $v$-$D$ phase diagrams of SC1 (fig. S2E) and SC2 (fig. S2F) present a crescent-like shape. In addition, fig. S2D plots $R_{xx}$ for the QAH state, revealing a resistance minimum at $v = 1$.

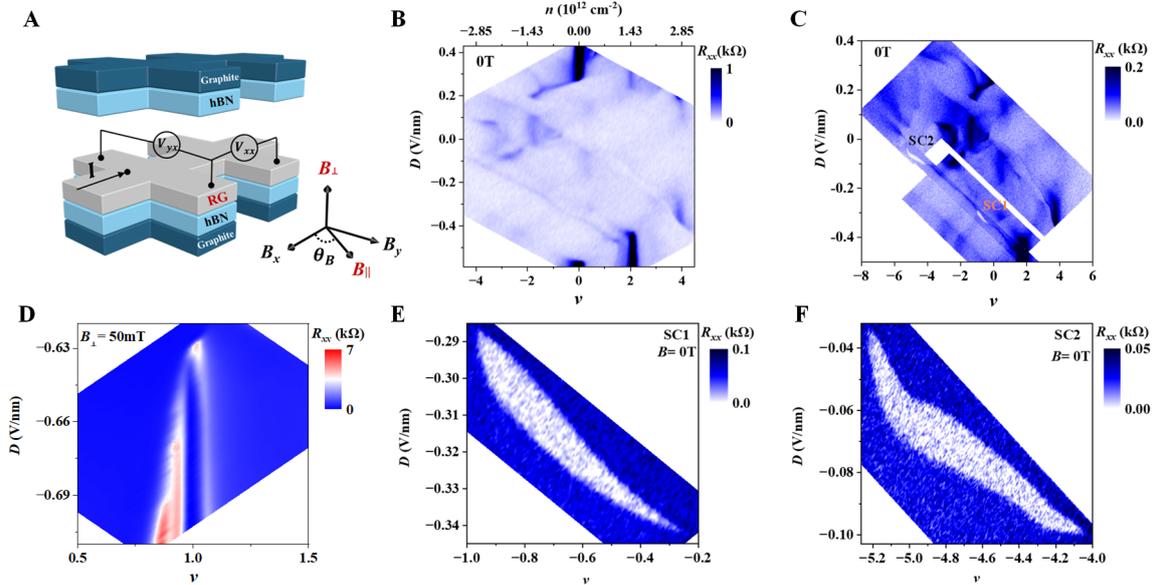

**Fig. S2. Measurement configuration and $v$-$D$ phase diagram. (A)** Schematic diagram of the transport measurement configuration. $B_∥$ and $B_⊥$ denote magnetic fields parallel and perpendicular to the two-dimensional plane, respectively. $B_∥$ is generated by the magnetic field components $B_x$ and $B_y$, and its in-plane rotation angle is denoted as $θ_B$. **(B and C)** Color maps of $R_{xx}$ as a function of carrier density $n$, filling factor $v$ and displacement field $D$ at zero magnetic field. (B) and (C) are the coarse and fine mappings, respectively. **(D)** Color plot of $R_{xx}$ as functions of $D$ and $v$ at $B_⊥ = 50$ mT. **(E and F)** Color maps of $R_{xx}$ as a function of $v$ and $D$ at 0 T for SC1 (E) and SC2 (F).

## Supplementary Note 3. Evolution of the Landau fan diagrams with *D*

The evolution of the Landau fan diagrams with *D* is illustrated in fig. S3. Two sets of QAH states with Chern numbers $C = +1$ and $C = -1$ are observed. The $C = +1$ state at positive $B_\perp$ gradually vanishes as *D* decreases, leaving only the $C = -1$ state.

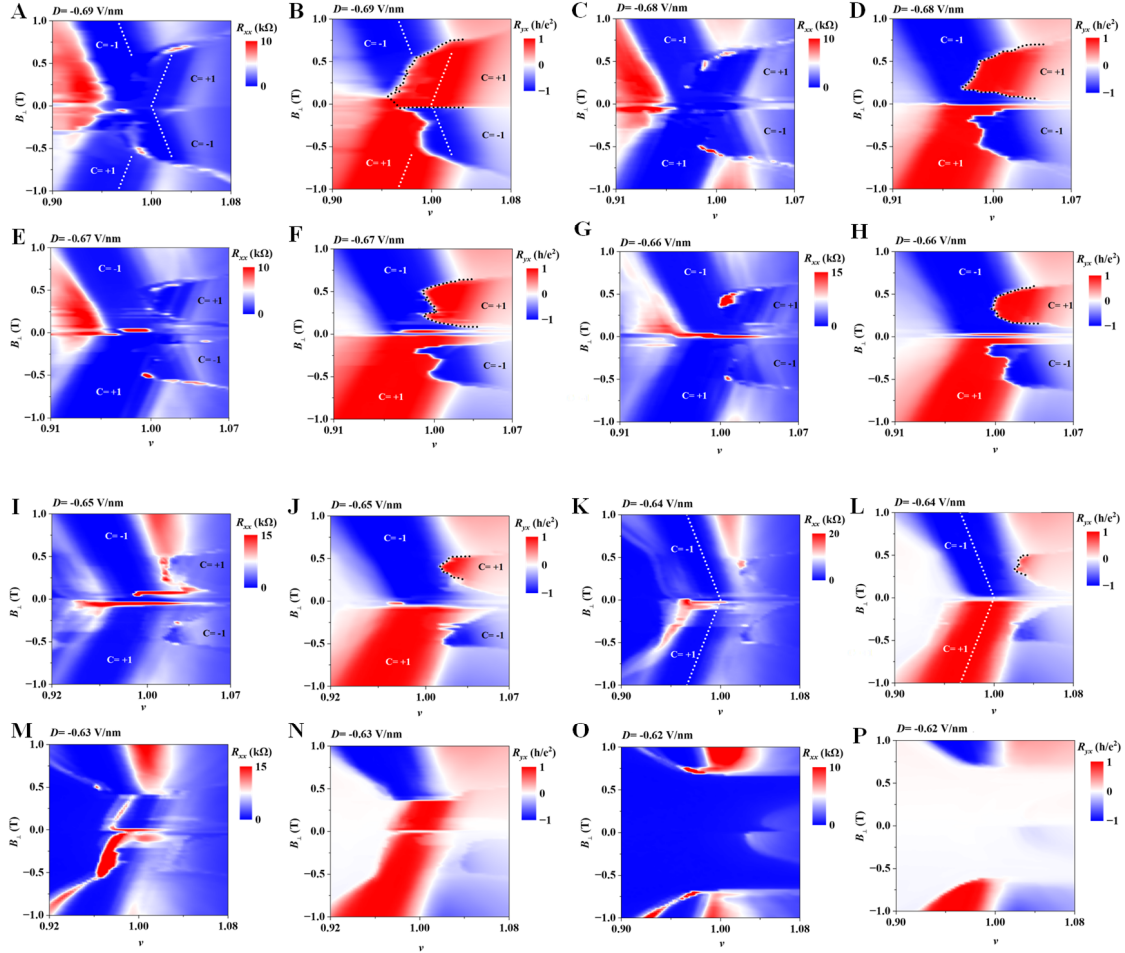

**Fig. S3. Landau fan diagrams under different *D*.** (**A** to **P**) Landau fan diagrams of $R_{xx}$ and $R_{yx}$ under $B_\parallel = 0$ T measured at $D = -0.69$ V/nm (A and B), -0.68 V/nm (C and D), -0.67 V/nm (E and F), -0.66 V/nm (G and H), -0.65 V/nm (I and J), -0.64 V/nm (K and L), -0.63 V/nm (M and N), and -0.62 V/nm (O and P), respectively. The black dashed lines in the $R_{yx}$ maps mark the critical positive $B_\perp$ ($B_{T\perp}$) for the phase transition from the $C = +1$ to the $C = -1$ QAH state.

## Supplementary Note 4. Evolution of the Landau fan diagrams with $B_\parallel$

As shown in fig. S4, the evolution of Landau fan diagrams under $D = -0.69$ V/nm with varying $B_\parallel$ illustrates the competition between two QAH states with Chern numbers $C = +1$ and $C = -1$. The $C = +1$ QAH state gradually weakens and eventually disappears with increasing $B_\parallel$, and only the $C = -1$ QAH state persists to $B_\parallel \geq 0.6$ T.

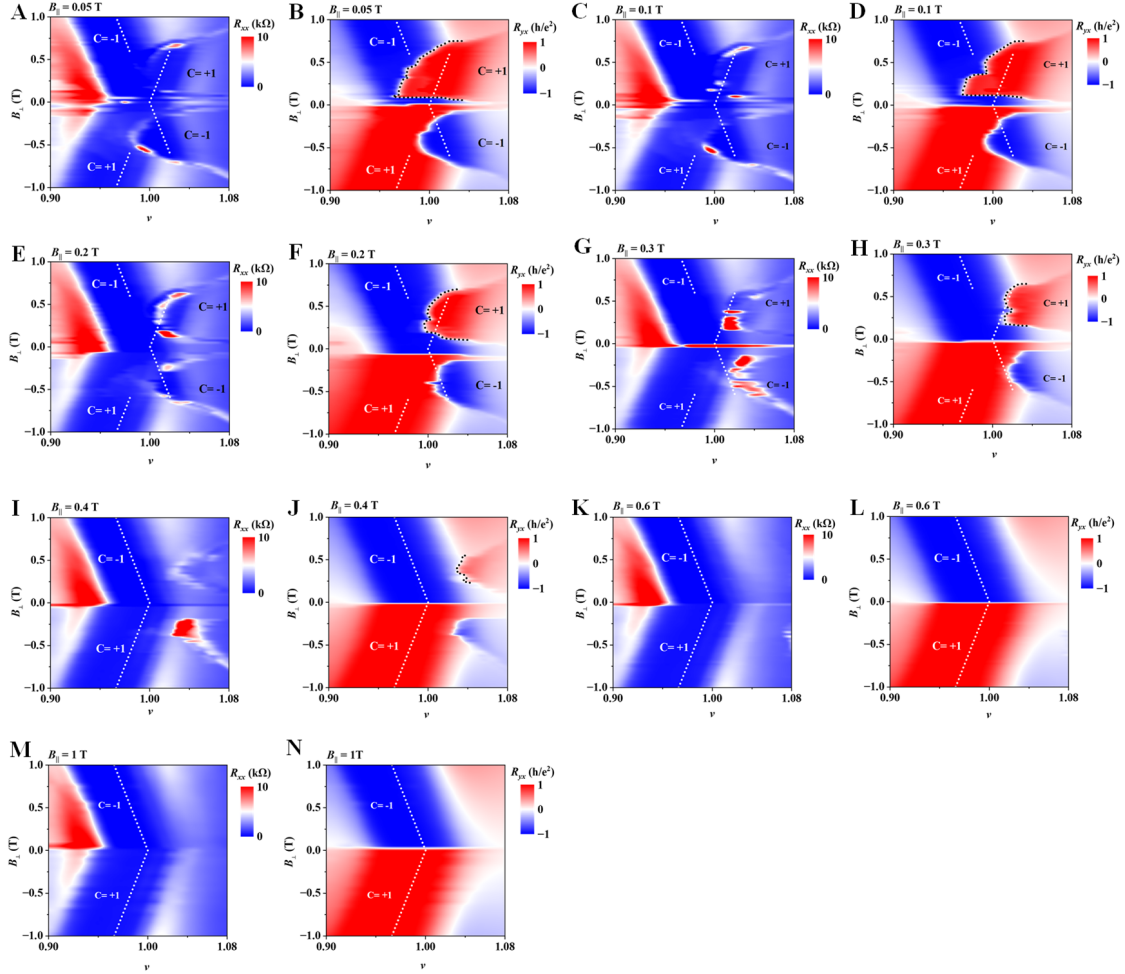

**Fig. S4. Landau fan diagrams under different $B_\parallel$.** (**A** to **N**) Landau fan diagrams of $R_{xx}$ and $R_{yx}$ under $D = -0.69$ V/nm measured at $B_\parallel = 0.05$ T (A and B), 0.1 T (C and D), 0.2 T (E and F), 0.3 T (G and H), 0.4 T (I and J), 0.6 T (K and L) and 1 T (M and N), respectively. White dashed lines show the expected evolution of Chern number $C = \pm 1$ QAH states based on the Streda formula $\partial n / \partial B_\perp = C\, e/h$. The black dashed lines in the $R_{yx}$ maps mark $B_{T\perp}$ for the transition from the $C = +1$ to the $C = -1$ QAH state.

## Supplementary Note 5. Isotropic response to in-plane magnetic field

We measured Landau fan diagrams at $D = -0.69$ V/nm at $B_\parallel = 0.4$ T (fig. S5) and 1 T (fig. S6) with various in-plane field angles $\theta_B$. The spectra acquired at different angles display negligible variations, demonstrating that the Landau fan response is highly isotropic to the in-plane magnetic field. At $D = -0.64$ V/nm, $B_\parallel$ also non-monotonically tunes the hysteretic window of the $C = -1$ QAH state, which first increases and then decreases with increasing $B_\parallel$ (fig. S7A). In contrast to $D = -0.69$ V/nm, no Chern number reversal from $C = +1$ to $C = -1$ occurs, and only the $C = -1$ state is maintained, as seen in the Landau fan diagrams from fig. S7, B to E. Similarly, we observe an isotropic response with respect to $\theta_B$ at $D = -0.64$ V/nm (fig. S7, F and G).

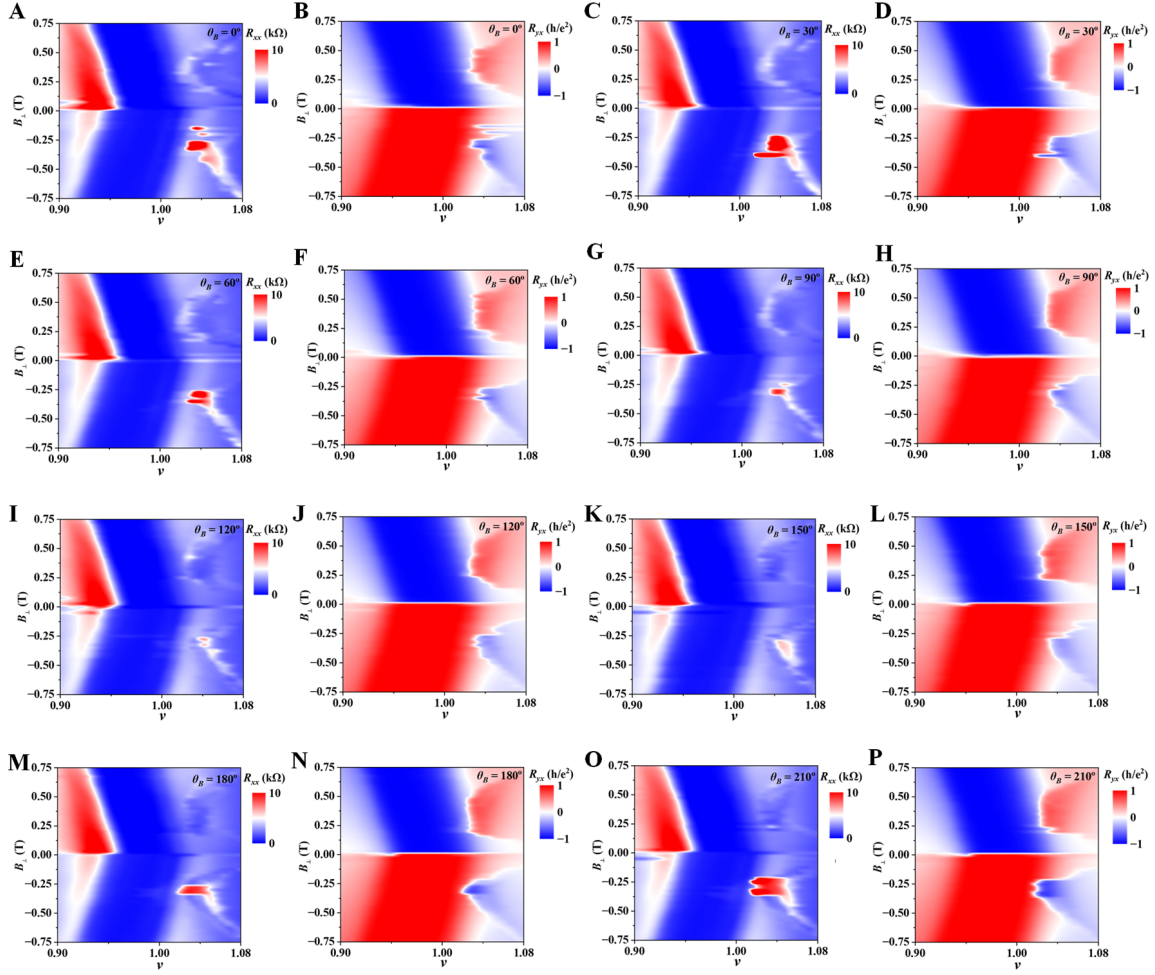

**Fig. S5. Landau fan diagrams at $B_\parallel$ = 0.4 T with different $\theta_B$.** (**A** to **P**) Landau fan diagrams of $R_{xx}$ and $R_{yx}$ under $D$ = -0.69 V/nm and $B_\parallel$ = 0.4 T measured with $\theta_B$ = 0° (A and B), 30° (C and D), 60° (E and F), 90° (G and H), 120° (I and J), 150° (K and L), 180° (M and N), and 210° (O and P), respectively.

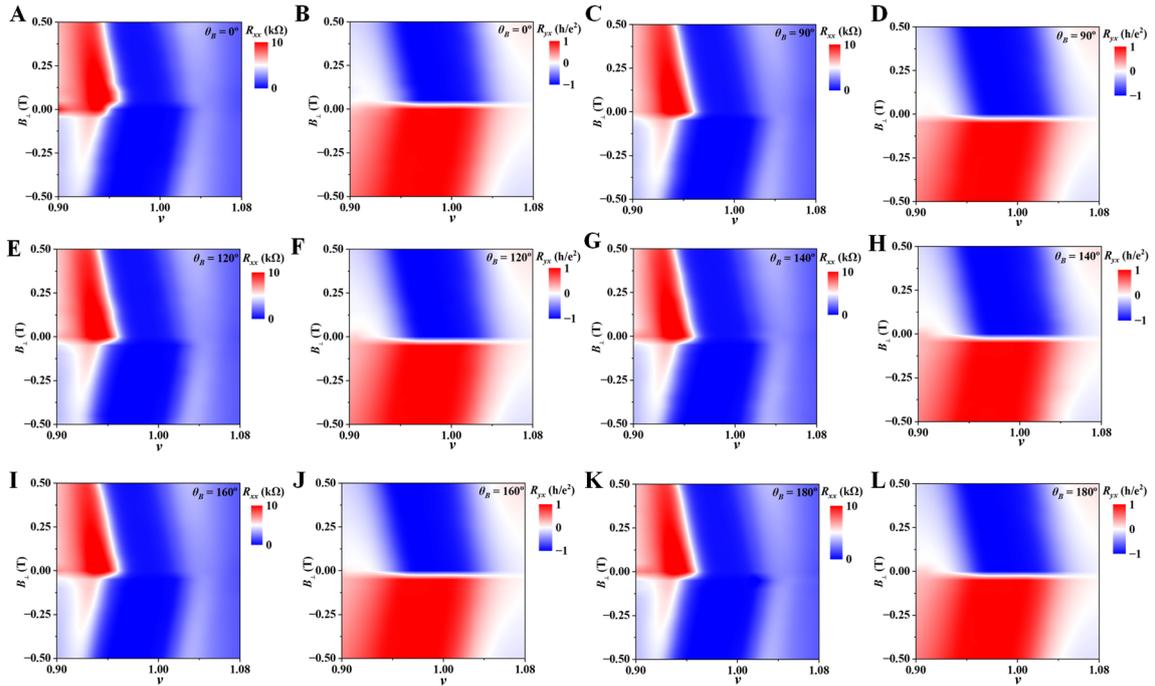

**Fig. S6. Landau fan diagrams at $B_{\parallel}$ = 1T with different $\theta_B$.** (A to L) Landau fan diagrams of $R_{xx}$ and $R_{yx}$ under $D$ = -0.69 V/nm and $B_{\parallel}$ = 1 T measured with $\theta_B$ = 0º (A and B), 90º (C and D), 120º (E and F), 140º (G and H), 160º (I and J), and 180º (K and L), respectively.

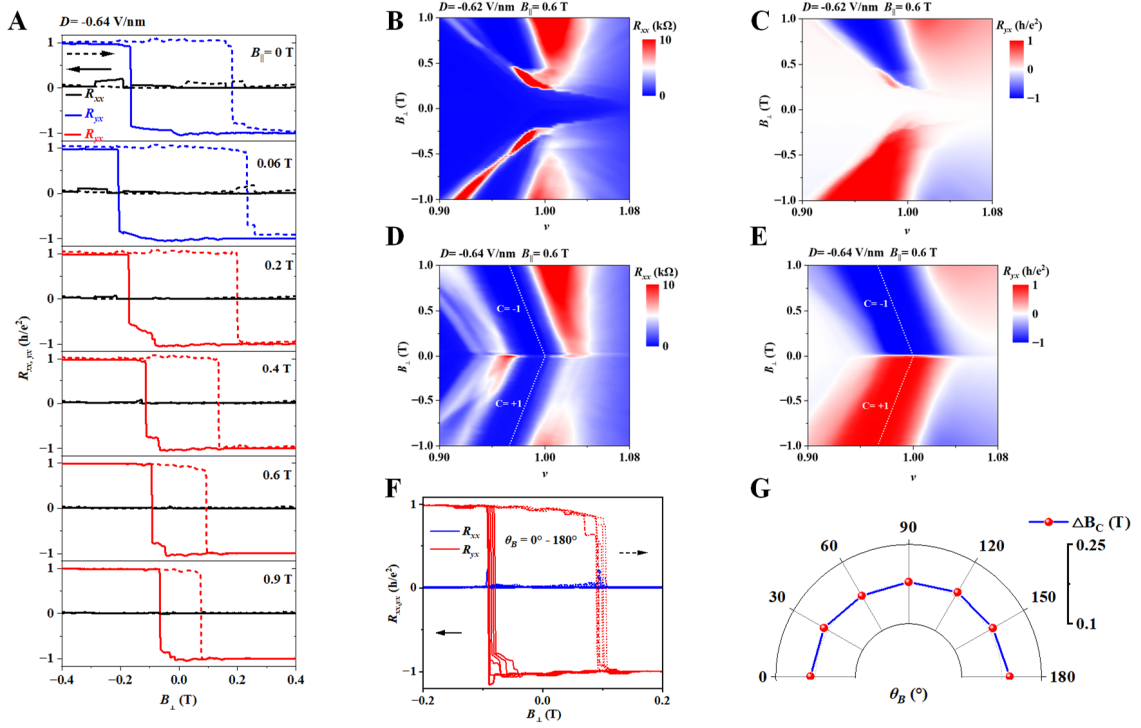

**Fig. S7. Modulation of QAH states by $B_{\parallel}$ at other $D$.** (A) Out-of-plane magnetic hysteresis loops measured under different $B_{\parallel}$ at $D$ = -0.64 V/nm and $v$ = 0.98. Solid (dashed) lines correspond to two different $B_{\perp}$ sweeping directions. For small $B_{\parallel}$ = 0 T, 0.06 T, 0.2 T, 0.4 T, 0.6 T and 0.9 T, the corresponding $B_C$ are ~ 0.17 T, ~ 0.22 T, ~ 0.19 T, ~ 0.12 T, ~ 0.09 T, and ~ 0.07 T, respectively. (B and C) Landau fan diagrams of $R_{xx}$ (B) and $R_{yx}$ (C) at $D$ = -0.62 V/nm measured under $B_{\parallel}$ = 0.6 T. (D and E) Landau fan diagrams of $R_{xx}$ (D) and $R_{yx}$ (E) at $D$ = -0.64 V/nm measured under $B_{\parallel}$ = 0.6 T. (F) Out-of-plane magnetic hysteresis loops measured at $D$ = -0.64 V/nm, $v$ = 0.98 and $B_{\parallel}$ = 0.6 T, under different $\theta_B$ = 0º-180º. (G) The out-of-plane magnetic hysteresis window $\Delta B_C$ of the QAH state was extracted from the hysteresis loops in fig. S7F for different $\theta_B$.

## Supplementary Note 6. Quantitative determination of the tilted $B_{\parallel}$ component

Any transport measurement under an in-plane magnetic field must eliminate the artifacts from its tilted component due to the unavoidable misalignment of $B_{\parallel}$ to the strictly in-plane direction. Here, we quantify this tilt using the d$V$/d$I$ map of the spin-polarized superconducting state in 8L-RG superlattices. From the measurements of d$V$/d$I$ versus $I_{dc}$ and $B_{\perp}$, the out-of-plane components for $B_y$ = 0.5 T, $B_y$ = -0.5 T, $B_x$ = 0.5 T, and $B_x$ = -0.5 T are -6, 8, 9, and -8 mT, respectively, extracted from the maxima of $I_{dc}$. The corresponding tilting angles are both below 1°, confirming that the measured $B_{\parallel}$ is an intrinsic physical property (fig. S8).

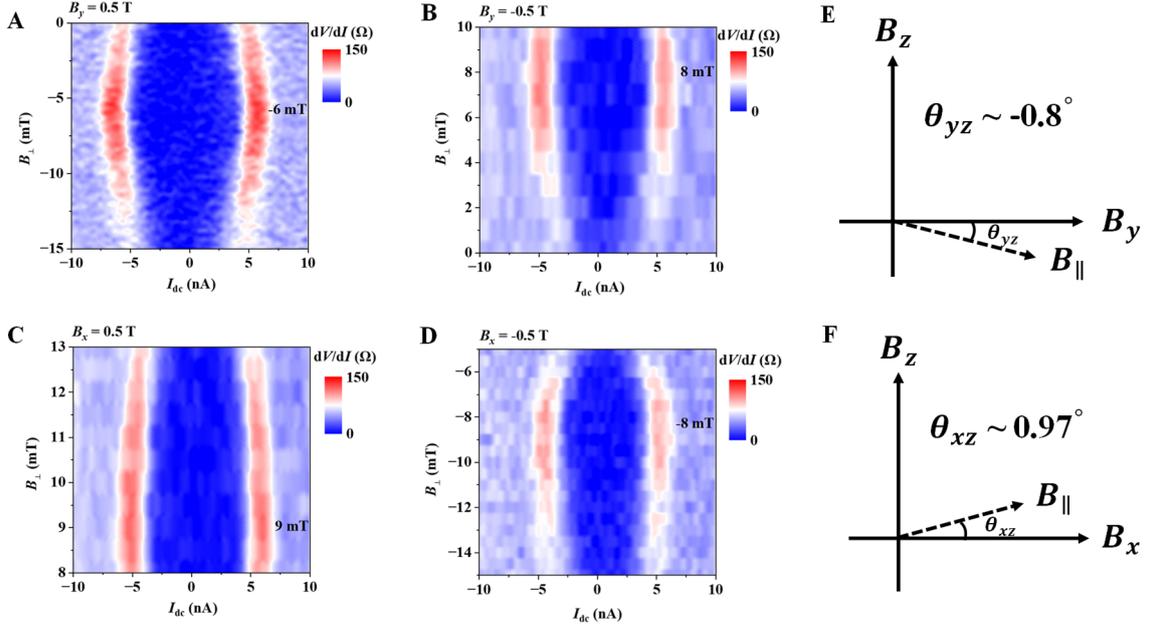

**Fig. S8. Quantitative estimation of the tilted component of $B_\parallel$.** (**A** to **D**) By measuring the d$V$/d$I$ maps in the spin-polarized superconducting SC3 state under $D$ = -0.19 V/nm and $v$ = -2.5 versus DC bias $I_{dc}$ and out-of-plane magnetic field $B_\perp$, the perpendicular components for $B_y$ = 0.5 T, $B_y$ = -0.5 T, $B_x$ = 0.5 T, and $B_x$ = -0.5 T were determined to be -6 mT, 8 mT, 9 mT, and -8 mT, respectively. These values were extracted from the positions of maximum critical current $I_{dc}$ in each map. (**E** and **F**) The tilting angle of the in-plane magnetic field along the $B_y$ axis is approximately -0.8°, and that along the $B_x$ axis is about 0.97°.

## Supplementary Note 7. The $B_\parallel$-modulated QAH state under various $B_\perp$

We measured the maps of $R_{xx}$ and $R_{yx}$ for the $B_\parallel$-modulated QAH state as functions of $v$ and $B_\parallel$ at $B_\perp$ = 0 T, 0.2 T, 0.4 T, 0.6 T, 0.7 T, 0.8 T, 0.9 T, -0.2 T, -0.4 T, and -0.6 T. We observed that the $C$ = +1 ($C$ = -1 at $B_\perp$ = -0.2 T, -0.4 T, and -0.6 T) state only emerges under a finite $B_\parallel$ and gradually vanishes with increasing $B_\perp$. The corresponding hysteresis also gradually deviates from quantization and disappears above $B_\perp$ > 0.5 T (fig. S9).

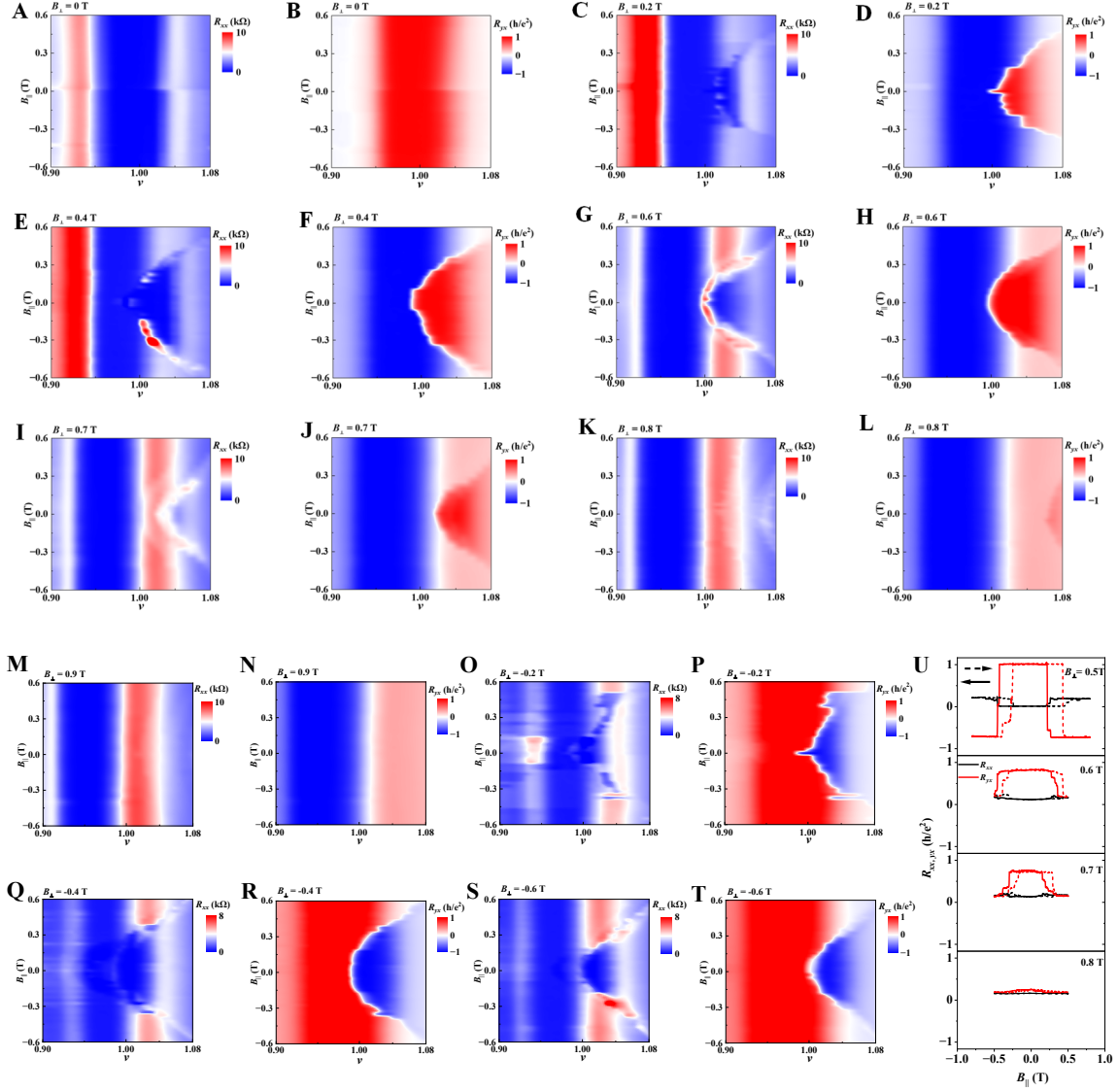

**Fig. S9. $B_\parallel$-modulated QAH state.** (**A** to **T**) Maps of $R_{xx}$ and $R_{yx}$ as functions of $\nu$ and $B_\parallel$ at $D$ = -0.69 V/nm, measured under $B_\perp$ = 0 T (A and B), 0.2 T (C and D), 0.4 T (E and F), 0.6 T (G and H), 0.7 T (I and J), 0.8 T (K and L), 0.9 T (M and N), -0.2 T (O and P), -0.4 T (Q and R), and -0.6 T (S and T). (**U**) In-plane magnetic hysteresis loops at $B_\perp$ = 0.5 T, 0.6 T, 0.7 T, and 0.8 T, measured at $D$ = -0.69 V/nm and $\nu$ = 1.01. Solid and dashed lines indicate different $B_\parallel$ sweeping directions.

## Supplementary Note 8. Theoretical calculations

### 8.1. The Hartree-Fock calculation

For each single-layer graphene, we choose the relative position of A and B sublattice as $\boldsymbol{\delta}_1 = \boldsymbol{\tau}_A - \boldsymbol{\tau}_B = -\frac{a_G}{\sqrt{3}}(0,1)$ in each layer where $a_G = 0.246$ nm is the graphene lattice constant. We label the top layer as the $0^{\text{th}}$ layer. The B sublattice in the $l$-th layer is chosen to be placed on top of the A sublattice in the $l + 1$-th layer. The lattice mismatch is $\epsilon_{latt} = (a_{hBN} - a_G)/a_G \approx 0.0167$, and the graphene on the top layer is aligned to hBN with twisting angle $\theta \approx 0.55°$. The Moiré reciprocal lattice vectors are then $\boldsymbol{b}_j = \boldsymbol{q}_{mod(j,3)+1} - \boldsymbol{q}_j$ where $\boldsymbol{q}_1 = \frac{4\pi}{3a_G}\left(1 - \frac{1}{1+\epsilon_{latt}}R(-\theta)\right)\hat{x}$ and $\boldsymbol{q}_j = R\left(\frac{2\pi(j-1)}{3}\right)\boldsymbol{q}_1$. A $D > 0$ displacement field points from the bottom layer to the top layer following the convention of the main text.

The single particle Hamiltonian in $K$ valley reads

$$H_K(\boldsymbol{r}) = H_{RLG,K}(-i\nabla_r) + V_{\text{moiré}}(\boldsymbol{r})$$

The rhombohedral L-layer graphene (RLG) part reads

$$H_{RLG,K}(\boldsymbol{k}) = \begin{pmatrix} v_F \boldsymbol{k}\cdot\boldsymbol{\sigma} & t^\dagger(\boldsymbol{k}) & t'^\dagger & \\ t(\boldsymbol{k}) & \ddots & \ddots & t'^\dagger \\ t' & \ddots & \ddots & t^\dagger(\boldsymbol{k}) \\ & t' & t(\boldsymbol{k}) & v_F \boldsymbol{k}\cdot\boldsymbol{\sigma} \end{pmatrix} + H_{ISP} + H_D$$

where $t(\boldsymbol{k}) = -\begin{pmatrix} v_4 k_+ & -t_1 \\ v_3 k_- & v_4 k_+ \end{pmatrix}, t' = \begin{pmatrix} 0 & 0 \\ t_2 & 0 \end{pmatrix}$ and $k_\pm = k_x \pm i k_y$. We have

$$H_{ISP} = V_{ISP}\left|l - \frac{L}{2}\right|\delta_{ll'}\delta_{aa'}, \quad H_D = -V_D(l - \frac{L}{2})\delta_{ll'}\delta_{aa'},$$

where $l, l'$ label the layer and $a, a'$ label the sublattice. $V_{ISP}$ is an inversion-symmetric potential binding electron to the rhombohedral graphene bulk, which is usually chosen to be zero. $V_D$ is the electrostatic potential from the displacement field and we choose the convention that $V_D > 0$ means $D > 0$. The Moiré coupling from the alignment with hBN reads

$$V_{\text{moiré}}(\boldsymbol{r}) = V_0 + \left[V_1 e^{i\psi_\xi}\sum_{j=1}^{3} e^{i\boldsymbol{b}_j\cdot\boldsymbol{r}}\begin{pmatrix} 1 & \omega^{-j} \\ \omega^{j+1} & \omega \end{pmatrix} + h.c.\right],$$

where $\psi_\xi$ = -136.55° and 16.55° for $\xi = 0,1$, which corresponds to the two types of stacking order where A,B sublattice of graphene is adjacent to the N,B atoms or B,N atoms of hBN, respectively. We choose $\xi = 0$ in our calculation. We use the parameters (6) as follows: $v_F = 630.63$ meV·nm, $v_3 = v_4 = 25.2$ meV·nm, $t_1 = 400$ meV, $t_2 = 0$, $V_0 = 0$, $V_1 = 20$ meV.

The interaction Hamiltonian we consider is

$$\hat{H}_I = \frac{1}{2\Omega_{tot}}\sum_{\boldsymbol{q}\in MBZ,G} V(\boldsymbol{q}):\rho_{\boldsymbol{q}}\rho_{-\boldsymbol{q}}:,$$

where $\rho_{\boldsymbol{q}} = \sum_{\boldsymbol{k}} c^\dagger_{\boldsymbol{k}+\boldsymbol{q},a,l,\eta,s} c_{\boldsymbol{k},a,l,\eta,s}$ and $a, l, \eta, s$ labels the sublattice, layer, valley and spin, respectively. $V(\boldsymbol{q})$ is the Fourier component of the electron-electron interaction, which we take to be the dual-gate-screened Coulomb interaction $V(\mathbf{q}) = \frac{e^2}{2\epsilon|q|}\tanh\frac{\xi|q|}{2}$.

$:\cdots:$ means normal-ordering the operators on certain reference state, which is determined by which interaction scheme is chosen to avoid the double counting of the interaction, as thoroughly discussed in previous work (7). We here use the charge neutrality point (CNP) scheme, which means that at CNP the mean field Hamiltonian of $\hat{H}_I$ is zero.

We then perform a self-consistent Hartree-Fock calculation on projected bands. We first diagonalize the single-particle Hamiltonian by $c^\dagger_{\mathbf{k+Q},a,l,\eta,s} = \sum_m u^{(\eta)}_{\mathbf{Q}al,m}(\mathbf{k}) c^\dagger_{\mathbf{k},m,,\eta,s}$ where $\mathbf{k}$ is restricted to the first Moiré Brillouin zone and $\mathbf{Q}$ take values in the Moiré reciprocal lattice. The total Hamiltonian is then projected to the active bands, which yields

$$\hat{H} = \sum_{m\eta s}\sum_{\mathbf{k}\in MBZ} \epsilon_{\mathbf{k}m\eta} c^\dagger_{\mathbf{k}m\eta s} c_{\mathbf{k}m\eta s}$$

$$+\frac{1}{2}\sum_{\substack{\mathbf{kk'q}\in MBZ}}\sum_{\substack{mm'nn'\\\eta\eta' ss'}} U^{\eta\eta'}_{mn,m'n'}(\mathbf{q};\mathbf{k},\mathbf{k'}):c^\dagger_{\mathbf{k+q}m\eta s}c_{\mathbf{k}n\eta s}c^\dagger_{\mathbf{k'-q}m'\eta's'}c_{\mathbf{k'}n'\eta's'}:$$

$U^{\eta\eta'}_{mn,m'n'}(\mathbf{q};\mathbf{k},\mathbf{k'})$ is the matrix element of the interaction on the active bands and $\epsilon_{\mathbf{k}m\eta}$ is the eigen energies of the single-particle Hamiltonian. Here, we project onto 3 valence bands and 7 conduction bands per valley per spin to study the $\nu = 1$ state, and we have verified that the results converge as the number of bands increases. By the standard Hartree-Fock decomposition, the interaction Hamiltonian is approximated by

$$\hat{H}^{MF}_I = \frac{1}{\Omega_{tot}}\sum_{\mathbf{k}\in fMBZ}\sum_{m\eta s,m'\eta's'} H^{MF}_{m\eta s,m'\eta's'}(\mathbf{k})\, c^\dagger_{\mathbf{k}m\eta s}c_{\mathbf{k}m'\eta's'}$$

where

$$H^{MF}_{m\eta s,m'\eta's'}(\mathbf{k})$$

$$= \sum_{nn'}\left(\delta_{ss'}\delta_{\eta\eta'}\sum_{\eta''s''} U^{(\eta\eta'')}_{mm',nn'}(0;\mathbf{k},\mathbf{k'})O_{n\eta''s'';n'\eta''s''}(\mathbf{k'})\right.$$

$$\left.- \sum_{\eta''} U^{(\eta\eta'')}_{mn',nm'}(\mathbf{k}-\mathbf{k'};\mathbf{k'},\mathbf{k})O_{n\eta's';n'\eta s}(\mathbf{k'})\right)$$

and the order parameter is $O_{m\eta s;m'\eta's'}(\mathbf{k}) = \langle c^\dagger_{\mathbf{k}m\eta s}c_{\mathbf{k}m'\eta's'}\rangle - O^{ref}_{m\eta s,m'\eta's'}(\mathbf{k})$. The reference order parameter in CNP scheme is chosen to be $O^{ref}_{m\eta s,m'\eta's'}(\mathbf{k}) =$

$\delta_{m \in \text{ valence band}} \delta_{m\eta s, m'\eta' s'}$ which let $O_{m\eta s; m'\eta' s'}(\mathbf{k})$ to be zero at the CNP of the non-interacting band. We then self-consistently determine the order parameter. We find that the ground state is a spin-valley-polarized insulator at $v = 1$ when $V_D = $ -30 meV, with $C$ = 1 in $K$ valley.

### 8.2. The competition between $C = -1$ and $C = +1$ states

The analysis here is similar to previous work (8) on rhombohedral hexalayer graphene. Because the effect of $B_\parallel$ on the QAH state is isotropic, we neglect the contribution from the in-plane orbital effect, which is generally anisotropic, and focus on the SOC $\lambda_I$ and the out-of-plane orbital magnetic moment $M_z$. Near $v = 1$, the hysteresis signals a symmetry-breaking state, which our Hartree-Fock calculation identifies as spin-valley polarized. The states with opposite valley polarization carry opposite Chern number due to time reversal symmetry. Therefore, the competition between $C = \pm 1$ states is indeed the competition between two valley polarized states. Hence, we consider the effective model, involving only the valley and spin degrees of freedom

$$H = -\frac{\lambda_I}{2}\tau_z s_z - B_\perp M_z \tau_z + \mu_B B_\perp s_z + \mu_B B_\parallel s_x. \quad (1)$$

Here, $\tau, s$ denotes the Pauli matrices acting on the valley and spin degrees of freedom respectively, with $\tau_z = +1(-1)$ corresponding to $K(K')$ valley, $s_z = +1(-1)$ corresponding to spin-up (down). We have already considered the electron $g$-factor ($g = 2$) and the spin-$\frac{1}{2}$ nature of the electron. We retain only the intrinsic Kane–Mele SOC (9). In the full basis, it is given by $\frac{\lambda_I}{2}\tau_z \sigma_z s_z$ where $\sigma_z$ denotes the Pauli matrix on sublattice degree of freedom. For $D < 0$, the conduction band mainly resides on the bottom-layer $B$ sublattice, such that $\langle \sigma_z \rangle \approx -1$, and the projected low-energy SOC reduces to the form used above and in the main text. The Rashba SOC, while symmetry-allowed since mirror symmetry is broken, is neglected because it is inter-sublattice and thus strongly suppressed upon projection to the relevant low-energy band. From the previous first principles calculation, we know that $\lambda_I > 0$ (8,10-12).

We first consider the case at $B_\parallel = 0$, where the Hamiltonian is diagonal. When $B_\perp > 0$, the flavor polarization $(\tau_z, s_z)$ of the ground state is

1. always $(-1, -1)$ if $M_z < 0$

2. $(-1, -1)$ when $B_\perp < \frac{\lambda_I}{2M_z}$ and $(+1, -1)$ when $B_\perp > \frac{\lambda_I}{2M_z}$ if $0 < M_z < \mu_B$.

3. $(+1, +1)$ when $B_\perp < \frac{\lambda_I}{2\mu_B}$ and $(+1, -1)$ when $B_\perp > \frac{\lambda_I}{2\mu_B}$ if $M_z > \mu_B$

Only in the second case, we can observe the Chern number reversal and the corresponding phase bound at $\frac{\lambda_I}{2M_z}$. This also means that the Chern band in $K$ valley should carry Chern number $C = -1$. This is opposite to the result of the Hartree-Fock calculation, while the same phenomena occur in previous experimental and theoretical study (8).

To study the effect of $B_\parallel$ we diagonalize (1). In general, $B_\parallel$ tends to tilt the spin towards the in-plane direction and suppress $s_z$, thereby reducing the effective strength of SOC and shifting the phase transition to a lower $B_\perp$. Given that $0 < M_z < \mu_B$, the lowest two states have energies $E_{\pm,+} = \pm B_\perp M_z \pm \mu_B\sqrt{(B \pm B_{SOC})^2 + B_x^2}$ where $B_{SOC} = \frac{\lambda_I}{2\mu_B}$ is the effective magnetic field per spin brought by SOC. The phase boundary is located where their energies are equal, which is

$$\left(\frac{B_\perp}{a}\right)^2 + \left(\frac{B_\parallel}{b}\right)^2 = 1 \quad (2)$$

where $a = \frac{\lambda_I}{2M_z} = \frac{\mu_B}{M_z}B_{SOC}$, $b = a\sqrt{1 - \left(\frac{M_z}{\mu_B}\right)^2}$. Therefore, we expect an elliptical phase boundary in the $(B_\parallel, B_\perp)$ plane, which is indeed supported by the experimental data in fig. S10A. From the lengths of the major and minor axes of the fitted phase boundary, we extract $\lambda_I$ and $M_z$ as functions of filling, as shown in fig.S10,B,C. We find that $\lambda_I$ stays around 60 μeV and is essentially independent of filling, as expected. Its magnitude is also consistent with previous experimental results ($\lambda_I \sim$ 40-120 μeV) in multilayer graphene (8,13-17).

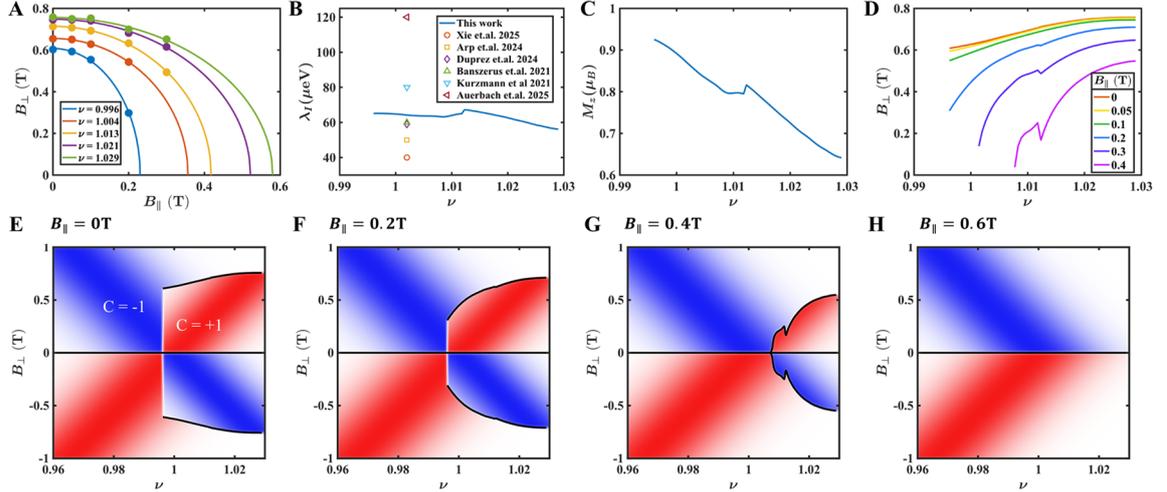

**Fig. S10. SOC strength and theoretically simulated Landau fans.** (**A**) The experimental values and the ellipse fitting curves of the phase boundary in ($B_\perp$, $B_\parallel$)-plane at different $\nu$. (**B** and **C**) The fitted SOC strength and orbital magnetic moment as a function of filling, obtained from the phase boundary in (A). We also compare our result with previous experiments (8,13-17) in (B). (**D**) The phase boundary in ($\nu$, $B_\perp$)-plane at different $B_\parallel$ calculated using the fitted SOC strength and orbital magnetic moment in (B) and (C). (**E** to **H**) Same phase boundaries as (D) but plotted separately for each $B_\parallel$, with schematic Landau fans overlaid. One should note that this is not a first-principle calculation of the full

phase diagram because we need to extract the SOC and orbital magnetic moment at each filling from experiment. The theory only predicts the $B_\parallel$ dependence of this phase diagram, i.e., the evolution from (E) to (H).

Using the extracted $\lambda_I$ and $M_z$ at each filling we plot the phase boundaries in $(\nu, B_\perp)$ plane under different $B_\parallel$ and in $(\nu, B_\parallel)$ plane at $B_\perp = 0.4$ T, as shown in fig. S10, D to H. One may note that since the Chern number reversal happens only when $0 < M_z < \mu_B$, the critical $B_\perp = \frac{\lambda_I}{2M_z}$ at $B_\parallel = 0$ is always larger than $B_{SOC} \approx 0.5$T in our analysis. We attribute the regions where the critical $B_\perp$ is smaller than $B_{SOC}$ to metastable states. This is also why no values of $\lambda_I$ or $M_z$ are extracted there. As the filling decreases, $M_z$ may exceed $\mu_B$, in which case the groundstate always has C = -1 when $B_\perp > 0$. The corresponding critical filling appears as a straight white line in fig. S12, D to G. This straight phase boundary is similar to the feature of the $B_\perp < 0$ part of Fig.1F.

### 8.3. The model for the magnetic hysteresis

#### 8.3.1. *Physical picture*

Here, we consider how the out-of-plane (in-plane) coercive field varies in the presence of an in-plane (out-of-plane) magnetic field, respectively. Since the state near $\nu \approx 1$ is a spontaneously symmetry-broken state in which spin and valley polarization are favored, both the spin and valley degrees of freedom can be described by Landau free energies with a Mexican-hat form. The total Landau free energy can be written as

$$F[\tau_z, s_z, s_x] = F_s[s] + F_v[\tau_z] + \mu_B B_\parallel s_x + \mu_B B_\perp s_z - \frac{\lambda_I}{2} s_z \tau_z - B_\perp M_z \tau_z \quad (3)$$

where $F_s[s], F_v[\tau_z]$ are potentials that favor the maximal spin and valley polarization, and $s^2 = s_x^2 + s_z^2 \leq 1, |\tau_z| \leq 1$.

The ground state minimizes the Landau free energy. However, the system may also be trapped in a local minimum rather than the global minimum, in which case a finite coercive field is required to drive the transition between them. We find that the qualitative behavior of the hysteresis does not depend on the detailed forms of $F_s[s]$ and $F_v[\tau_z]$. Instead, the essential ingredients are as follows.

We focus on the effective Landau free energy of the valley after the free energy of spin part is minimized, as the valley flipping is directly related to the transition of the Chern number. Before applying $B_\perp$, the Landau free energy is symmetric under $\tau_z \leftrightarrow -\tau_z$, as shown in the leftmost part of figs. S11C,D and S12C,D. A finite $B_\perp$ tilts the free energy curve leftward or rightward, depending on whether the effect of SOC or the out-of-plane orbital magnetism is dominant. We define a quantity $B_{\text{eff},v}$ ($v$ denotes 'valley') to describe how one valley is favored over another and $B_{\text{eff},v}$ must exceed a coercive field $H_{c,v}$ to induce valley switching. If the tunneling between the local minima can be neglected, we just define $B_{\text{eff},v}$ as $B_\perp$-induced effective magnetic field felt by valley:

$$B_{\text{eff},v} = -\frac{M_z}{\mu_B}B_\perp - \frac{\lambda_I}{2\mu_B}s_z. \quad (4)$$

which satisifies $F[\tau_z, s_z, s_x] = F_v[\tau_z] + \mu_B B_{\text{eff},v}\tau_z + (\tau_z \text{ independent terms})$.

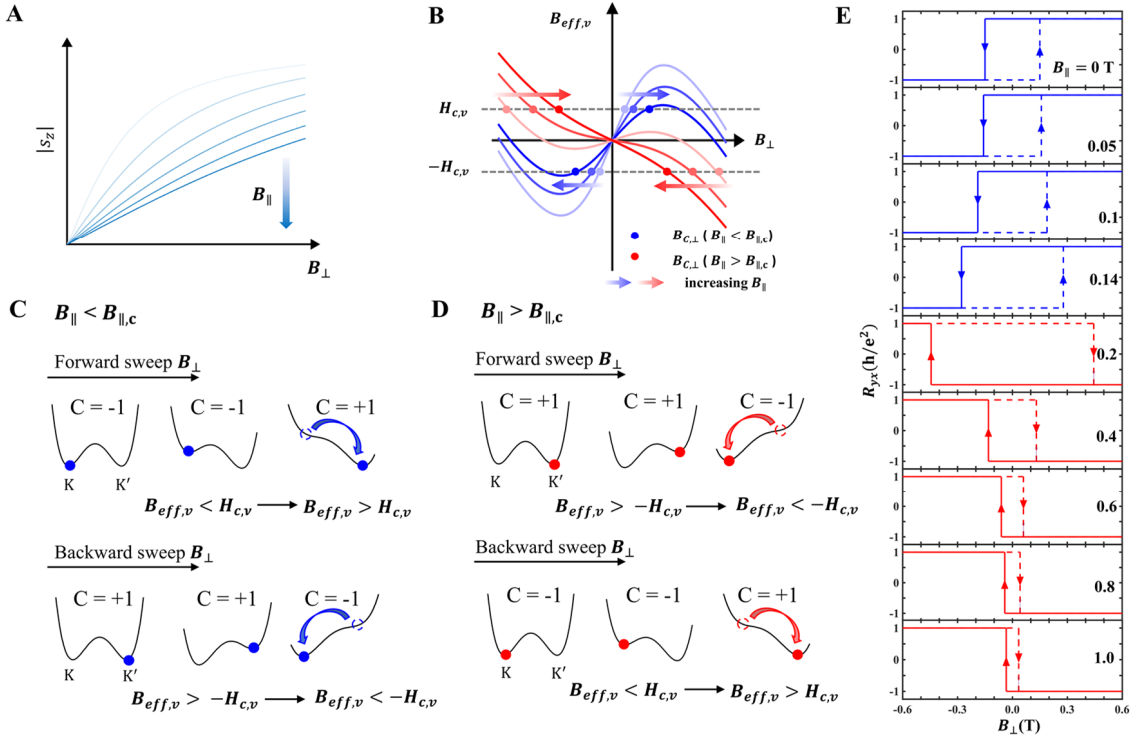

**Fig. S11. Critical field and hysteresis evolution under out-of-plane magnetic field.** (**A**) Schematic $|s_z(B_\perp)|$ for different $B_\parallel$. (**B**) Sketch of the effect of $B_\parallel$ on $B_{eff,v}$ and the evolution of $B_{C,\perp}$. $B_{eff,v}$ is the effective magnetic field acting on the valley degree of freedom, including the contribution from both the orbital magnetic moments and SOC. In (A and B), $B_\parallel$ increases from light to dark. Circles in (B) mark the Chern-number reversal, and the arrow shows the shift of $B_c$ with increasing $B_\parallel$. The blue and red curves/circles in (B to D) correspond to $B_\parallel < B_{\parallel,c}$ and $B_\parallel > B_{\parallel,c}$, respectively. $B_{\parallel,c}$ is defined as the critical value of $B_\parallel$ above which $B_{eff,v} < H_{c,v}$ for all $B_\perp > 0$, which is the critical $B_\parallel$ where $C = +1$ state (when $B_\perp > 0$) vanishes in the phase diagram. (**C and D**) The sketch of Landau free energy of valley degree of freedom and their evolution under $B_\perp$ when $B_\parallel < B_{\parallel,c}$ and $B_\parallel > B_{\parallel,c}$, respectively. $H_{c,v}$ and $-H_{c,v}$ always correspond to the transition from $C = -1$ to $C = +1$ and from $C = +1$ to $C = -1$, respectively. (**E**) The simulation of the hysteresis when sweeping $B_\perp$ at different $B_\parallel$. Solid (dashed) lines correspond to sweeping $B_\perp$ from positive (negative) values to negative (positive) values. See Supplementary Note 8.3 for detailed explanation.

Through SOC, $s_z$ generates an effective field acting on the valley sector, which drives the switching of valley polarization and the corresponding Chern number. Combining this effect with the orbital magnetic moment term $-B_\perp M_z\tau_z$.

Since tunneling may lead to the fact that the transition happens before the local minimum becomes local maximum, to mimic in the following, we will also define $B_{\text{eff},v}$ by letting $\mu_B B_{\text{eff},v}$ to be the energy difference between the local and global minima and treat $H_{c,v}$ as a fixed phenonological parameter, thereby incorporating the effect of tunneling across the barrier. These two definitions of $B_{\text{eff},v}$ have a similar dependence on $B_\perp$ and $B_\parallel$. The representative schematic curves are shown in figs. S11B and S12B.

At zero $B_\parallel$, when $B_\perp$ is small, $B_{\text{eff},v}$ initially increases with $B_\perp$, because $-s_z$ rises rapidly from 0 (or -1) to 1 (notice that spin is always anti-parallel to magnetic field) (fig. S11A). The SOC effect will be more important and $C = 1$ state is more favorable. When $B_{\text{eff},v}$ reaches the critical value $H_{c,v}$ (blue points in fig. S11B), it is strong enough to drive the system from the local minimum with $C = -1$ to the global minimum with $C = +1$, as depicted in fig. S11C. Once $|s_z|$ saturates at 1, however, the orbital term $-(M_z/\mu_B)B_\perp$ dominates, and the slope of $B_{\text{eff},v}$ becomes negative, which means that $C = -1$ state is further favored, and the switching from $C = +1$ to $C = -1$ happens when $B_{\text{eff},v} = -H_{c,v}$.

We then consider the evolution of the $B_\perp$-induced hysteresis under $B_\parallel$, which suppresses $s_z$ (fig. S11A) and shifts $B_{\text{eff},v}$ toward more negative values ($C = +1$ less favorable) as $B_\parallel$ increases (fig. S11B). Consequently, for the $C = -1$ to $C = +1$ transition, which requires a positive $B_{\text{eff},v}$, the critical $B_\perp$ increases with $B_\parallel$, as marked by the blue points in fig. S11B. In contrast, for the $C = +1$ to $C = -1$ transition, which requires a negative $B_{\text{eff},v}$, the critical $B_\perp$ decreases, as marked by the blue points in fig. S11B. Indeed, this can be simply understood by the fact that the $C = -1$ region which is disfavored by SOC will always expand when increasing $B_\parallel$ suppresses SOC, as discussed in the main text. When $B_\parallel$ is larger than a certain $B_{\parallel,c}$, $B_{\text{eff},v}$ is always smaller than $H_{c,v}$ and there is no $C = +1$ groundstate at $B_\perp > 0$. Instead, when $B_\perp$ is scanned from a negative value, where the system should be a $C = +1$ state for large enough $|B_\perp|$, the system will remain in this $C = +1$ local minimum till $B_{\text{eff},v} < -H_{c,v}$, as shown by fig. S11D and the corresponding critical $B_\perp$ is marked by the red points in fig. S11B.

A similar analysis also applies to the effect of $B_\perp$ on the critical $B_\parallel$. Applying a $B_\perp$ may either enhance $B_{\text{eff},v}$ or suppress $B_{\text{eff},v}$, depending on whether the out-of-plane orbital magnetism, or the increasing of spin magnetism is dominant. As a result, the $B_{\text{eff},v}$ may shift either upaward or downaward as shown in fig. S12B, while it is a monotonic function of $B_\parallel$, which solely suppresses the effect of spin magnetism. When $B_\perp < B_{T\perp}^{(0)}$, maximum of $B_{\text{eff},v}$ is larger than $H_{c,v}$ so the spin magnetism induced field by SOC is strong enough to drive the state to $C = +1$ favored by SOC. The switching between $C = \pm 1$ also happens when $|B_{\text{eff},v}| = H_{c,v}$ as marked by the solid and empty red circles in fig. S12B and the corresponding changes of the Landau energy are shown in fig. S12 C. Whereas, when $B_\perp > B_{T\perp}^{(0)}$ and $B_{\text{eff},v}$ is always smaller than $H_{c,v}$, the system always stay at $C = -1$.

### 8.3.2. *Detailed Model*

Given the overall picture above, we now present more detailed analysis. We first consider the $B_\parallel = 0$ case. To minimize $F[\tau_z, s_z, s_x]$, one has $s_z = \pm 1$ so $s_x$ is always zero. The free energy then reduces to

$$F[\tau_z, s_z] = F_s[|s_z|] + F_v[\tau_z] + \mu_B B_\perp s_z - \frac{\lambda_I}{2} s_z \tau_z - B_\perp M_z \tau_z \quad (5)$$

If thermal fluctuations and quantum tunneling effects are both negligible, the hysteresis is determined by the point at which the local minimum loses stability and becomes an unstable extremum. We begin with the local minimum at $C = -1$. Experimentally, the Hall conductance remains quantized before the transition to $C = +1$, indicating that the system remains fully valley polarized prior to the switching. In this case, the free energy becomes

$$F[\tau_z = 1, s_z] = F_s[|s_z|] + (\mu_B B_\perp - \frac{\lambda_I}{2}) s_z + const \quad (6)$$

where the constant part means independent of $s_z$. Therefore, $B_\perp$ must be at least larger than $\frac{\lambda_I}{2\mu_B} \approx 0.5\text{T}$ to drive the spin from up to down. This is in clear contrast to the experiment, where $B_\perp \approx 0.1\,\text{T}$ is already sufficient to flip the Chern number. We therefore conclude that thermal fluctuations and quantum tunneling are important, and that the system switches between different local minima before the original local minimum becomes unstable.

To model this effect, we adopt a simple phenomenological criterion. We assume a critical free-energy difference $\Delta F_c$: when the free-energy difference $\Delta F$ between the current local minimum and the global minimum reaches $\Delta F_c$, the system switches to the global minimum. We then define an effective magnetic field acting on the valley degree of freedom, $B_{\text{eff},v}$, through $\Delta F = 2\mu_B B_{\text{eff},v}$, and denote the corresponding critical field by $H_{c,v}$. This gives

$$2\mu_B B_{eff,v} = \min_{s_z} F[\tau_z = 1, s_z] - \min_{s_z} F[\tau_z = -1, s_z]$$

$$= -\left|\mu_B B_\perp - \frac{\lambda_I}{2}\right| + \left(\mu_B B_\perp + \frac{\lambda_I}{2}\right) - 2M_z B_\perp \quad (7)$$

namely

$$B_{eff,v} = \begin{cases} \left(1 - \dfrac{M_z}{\mu_B}\right) B_\perp, & B_\perp \leq B_{SOC} \\ B_{SOC} - \dfrac{M_z}{\mu_B} B_\perp, & B_\perp > B_{SOC} \end{cases} \quad (8)$$

where $B_{SOC} = \frac{\lambda_I}{2\mu_B}$.

When an in-plane magnetic field is applied, we need to minimize the full free energy $F[\tau_z, s_z, s_x]$. We still assume $\tau_z = \pm 1$, so that

$$F[\tau_z = \pm 1, s_z, s_x] = F_s[s] + F_v[\pm 1] + \mu_B B_\| s_x + \mu_B(B_\perp \mp B_{SOC})s_z \mp B_\perp M_z$$

By time reversal symmetry $F_v[1] = F_v[-1]$. Minimizing the free energy, the spin components take the form

$$s_{z,\pm} = -\frac{B_{\perp,eff,\pm}}{\sqrt{B_{\perp,eff,\pm}^2 + B_\|^2}}, B_{\perp,eff,\pm} = B_\perp \mp B_{SOC}, s_{x,\pm} = -\sqrt{1 - s_{z,\pm}^2}, \quad (9)$$

and the corresponding free energy is

$$F[\tau_z = \pm 1] = F_s[1] + F_v[1] - \mu_B\sqrt{(B_\perp \mp B_{SOC})^2 + B_\|^2} \mp B_\perp M_z. \quad (10)$$

We then have

$$B_{eff,v} = \frac{1}{2}\left(\sqrt{(B_\perp + B_{SOC})^2 + B_\|^2} - \sqrt{(B_\perp - B_{SOC})^2 + B_\|^2}\right) - \frac{M_z}{\mu_B}B_\perp, \quad (11)$$

which behave like figs. S11B and S12B. The transition points in $B_\perp$ at fixed $B_\|$, and in $B_\|$ at fixed $B_\perp$, can then be determined from the condition $|B_{eff,v}| = H_{c,v}$. Using the parameter $\lambda_I = 60\ \mu eV, M_z = 0.9\ \mu_B, H_{c,v} = 0.015$ T, we obtain figs. S11E and S12E, which captures the main features of the experimental results in Fig. 2A and H.

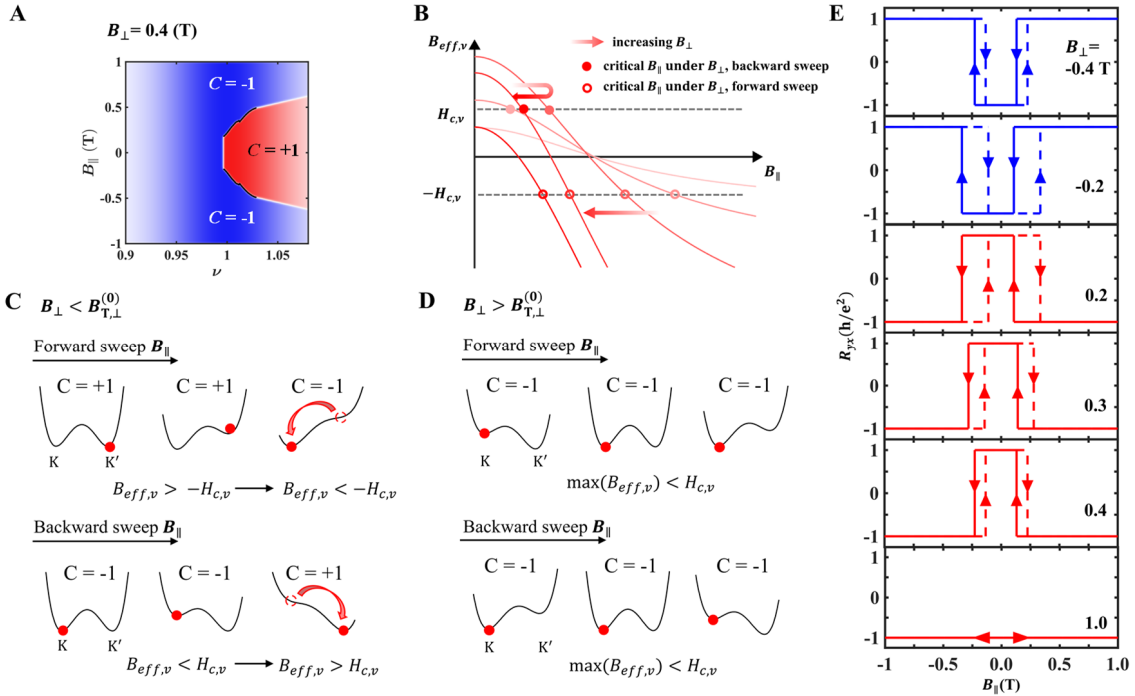

**Fig. S12. Phase boundary, critical Field and hysteresis rvolution under in-plane magnetic field.** (**A**) The phase boundary in $(v, B_\|)$ plane at $B_z = 0.4$ T, calculated using the fitted SOC strength and orbital magnetism. (**B**) Sketch of the effect of $B_\perp$ on $B_{eff,v}$ and the evolution of critical $B_\|$. In (B), $B_\|$ increases from light to dark. Circles in (B) mark the Chern-number reversal, and the arrow shows

the shift of critical $B_\parallel$ with increasing $B_\perp$. The evolution of the critical $B_\parallel$ may be non-monotonic, which is the result of the non-monotonic dependent of $B_{eff,v}$ in $B_\perp$ when fixing $B_\parallel$. (**C** and **D**) The sketch of Landau free energy of valley degree of freedom and their evolution under $B_\parallel$ at different $B_\perp$ with $B_\perp < B_{T\perp}^{(0)}$ and $B_\perp > B_{T\perp}^{(0)}$, respectively. Here $B_{T\perp}^{(0)}$ is defined as the critical out of plane magnetic field of the chirality switching ($B_{T\perp}$) at $B_\parallel = 0$. (**E**) The simulation of the hysteresis when sweeping $B_\perp$ at different $B_\parallel$. Solid (dashed) lines indicate different $B_\parallel$ sweeping directions. See Supplementary Note 8.3 for detailed explanation.

### 8.4. Analysis of the SC1 and SC3 phases

Here, we discuss the possible pairing spin structure of SC1 and SC3. To lowest order, the symmetry-allowed free energy involving both spin-singlet and spin-triplet pairing order parameters can be written as

$$F = \sum_{s_z} A_{t,s_z}\Phi^\dagger_{t,s_z}\Phi_{t,s_z} + \sum_{s_z} A_s \Phi^\dagger_s \Phi_s + \sum_{i=x,y,z}\sum_{s_z,s_z'} g\mu_B B_i \cdot \Phi^\dagger_{t,s_z}[S_i]_{s_z,s_z'}\Phi_{t,s_z'}$$

$$+\Lambda(\Phi^\dagger_s \Phi_{t,0} + h.c.)$$

where $\Phi_{t,s_z}$ denotes the $s_z$ component of the spin-triplet pairing order parameter, $s_z = +1$ for $\uparrow\uparrow$, $s_z = 0$ for $(\uparrow\downarrow + \downarrow\uparrow)/\sqrt{2}$ $s_z = -1$ for $\downarrow\downarrow$, while $\Phi_s$ denotes the spin-singlet pairing order parameter. $S_i$ are the spin-1 matrices with $S_z = \mathrm{diag}(1,0,-1)$, $[S_x \pm iS_y]_{s_z,s_z'} = \sqrt{2}\delta_{s_z,s_z'+1}$, and $g$ is the Lande g-factor.

In the absence of both SOC and magnetic field, SU(2) spin-rotation symmetry requires $\Lambda = 0$ and $A_{t,+1} = A_{t,0} = A_{t,-1}$. In the absence of in-plane magnetic field, the Ising SOC breaks the full SU(2) rotation symmetry down to the residual U(1) spin-rotation symmetry about the $z$ axis. As a result, a finite $\Lambda$ is allowed, whereas mixing between $\Phi_s$ and the $s_z = \pm 1$ component of $\Phi_t$, as well as the mixing between different triplet components, remains forbidden. The coefficient $\Lambda$ is at most of the order of the SOC strength. Furthermore, time reversal symmetry requires $A_{t,+1} = A_{t,-1}$. The sign of $A$ determines whether a superconducting instability develops in the corresponding channel.

Given that SC1 is slightly enhanced and SC3 is induced by an in-plane magnetic field, it is unlikely that they are solely spin-singlet pairing. There are two possibilities: a mixture of spin singlet and spin triplet, or only the spin-triplet pairing.

The mixing between spin singlet and spin triplet can exist only in the presence of SOC. The relevant comparison is between the SOC strength and the difference between $A_{t,s_z}$ and $A_s$ instead of the SOC strength and $T_c$. The difference between $A_{t,s_z}$ and $A_s$ is typically large because it is the splitting between microscopic interactions in the singlet and triplet channel and, although its precise magnitude depends on the pairing mechanism, it is generally not expected to be as small as the SOC scale. In the special case where both $\Phi_t$ and $\Phi_s$ correspond to intervalley pairing and the system preserves $SU(2)_+ \times SU(2)_-$ spin-rotation symmetry in the two valleys separately, one has $A_t = A_s$. However, in rhombohedral multilayer graphene system, this symmetry is usually broken by, for example, Hund's rule coupling (*17-20*) or electron-phonon-mediated interactions (*21-22*), both of which are typically estimated to be on the order of several meV. This is much larger than the intrinsic SOC strength (~60 μeV) in this system (without proximity to a substrate with strong SOC like WSe₂), yielding a weak singlet-triplet mixing.

We next turn to the case of spin-triplet pairing, which can be naturally stabilized by an in-plane magnetic field. Since the in-plane orbital effect is already shown to be negligible by isotropic response of the in-plane magnetic field, the primary effect of the field is to couple to spin, allowing the spin-triplet pairing state to gain Zeeman energy. Specifically, for an in-plane magnetic field applied along the $x$ direction without losing generality, the free energy is given by

$$F = \Phi_t^\dagger \begin{pmatrix} A_{t,1} & \frac{1}{\sqrt{2}} g\mu_B B_x & 0 \\ \frac{1}{\sqrt{2}} g\mu_B B_x & A_{t,0} & \frac{1}{\sqrt{2}} g\mu_B B_x \\ 0 & \frac{1}{\sqrt{2}} g\mu_B B_x & A_{t,1} \end{pmatrix} \Phi_t$$

where $\Phi_t = (\Phi_{t,1}, \Phi_{t,0}, \Phi_{t,-1})$.

The smallest eigenvalue of the mass matrix is then $A = \frac{A_{t,0}+A_{t,1}}{2} - \sqrt{\left(\frac{A_{t,0}-A_{t,1}}{2}\right)^2 + (g\mu_B B_x)^2}$. In this case, if $\min\{A_{t,1}, A_{t,0}\}$ is already negative, then $A < 0$ at $B_x = 0$. The system is therefore superconducting at zero magnetic field and can be further stabilized by a finite $B_x$, which may correspond to the case of SC1. At $B_x = 0$, the pairing is either the ↑↑/↓↓ component or the ↑↓+↓↑ component, depending on whether $A_{t,1}$ or $A_{t,0}$ is smaller. If $\min\{A_{t,1}, A_{t,0}\}$ is positive but small, a finite $B_x$ may drive $A$ from positive to negative. In that case, the system is non-superconducting at zero field but can be tuned to a superconducting phase by applying $B_x$, which may correspond to SC3.

The splitting between $A_{t,1}$ and $A_{t,0}$, if exists, must come from SOC that breaks SU(2). The nesting condition between TRS-related Fermi surfaces would favor superconductivity in the $s_z = 0$ channel. However, there may be many other possibilities, and the actual pairing symmetry of the superconducting states in 8L-RG is yet to be explored in future experiments.

**Supplementary Note 9. The experimental results on 8L-RG Device D2**

Fig. S13 presents the experimental data from another 8L-RG Device D2 to demonstrate that the main results from Device D1 are reproducible. From $R_{xx}$ features at carrier densities $n \approx 2.85 \times 10^{12}$ cm$^{-2}$ and $n \approx 1.43 \times 10^{12}$ cm$^{-2}$ in fig. S13A, we determine a similar twist angle of $\theta \approx 0.55°$ between the 8L-RG and hBN (*1-3*).

For hole-doped carriers near the moiré superlattice, we also observe the superconducting states SC1 and SC2 at zero magnetic field and the $B_\parallel$-induced superconducting state SC3 under $B_\parallel = 0.5$ T, as shown in fig. S13, B and C. Meanwhile, in the electron-doped region far from the moiré superlattice, we observe the QAH state (fig. S13, D to F). The Chern number reversal in the QAH state can be tuned by $D$ and $B_\parallel$, corresponding to a topological transition between Chern number $C = +1$ and $C = -1$ (fig. S13, G and H, K and M). The $B_\parallel$ can non-monotonically tune the hysteretic window and chirality reversal of the QAH state, and the QAH state shows an isotropic response to $B_\parallel$ (fig. S13, L and M). Finally, we also observe quantized hysteresis modulated by $B_\parallel$ (fig. S13, I and J).

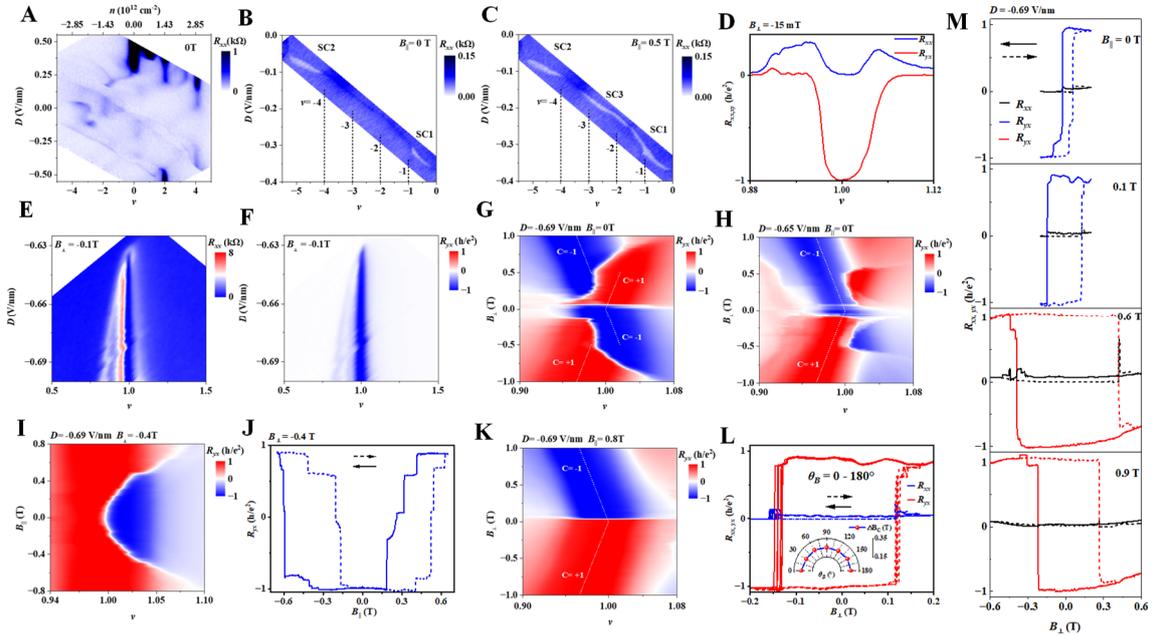

**Fig. S13. Reproducible measurements in Device D2.** (**A** to **C**) Color maps of $R_{xx}$ as a function of $n$, $v$ and $D$, acquired under coarse scan at 0 T (A), fine scan at 0 T (B), and fine scan with $B_\parallel = 0.5$ T (C), respectively. (**D**) The curves of $R_{xx}$ and $R_{yx}$ as a function of $v$ are obtained at $D = -0.69$ V/nm and $B_\perp = -15$ mT. (**E** and **F**) Color plots of the $R_{xx}$ (E) and $R_{yx}$ (F) as functions of $D$ and $v$ at $B_\perp = -100$ mT. (**G** and **H**) Landau fan diagrams of $R_{yx}$ under $B_\parallel = 0$ T measured at $D = -0.69$ V/nm (G) and $D = -0.65$ V/nm (H), respectively. White dashed lines show the expected evolution of Chern number $C = \pm 1$ QAH states based on the Streda formula $\partial n / \partial B_\perp = C\, e/h$. (**I**) Maps of $R_{yx}$ as functions of $v$ and $B_\parallel$ at $D = -0.69$ V/nm, measured under $B_\perp = -0.4$ T. (**J**) In-plane magnetic hysteresis loop at $B_\perp = -0.4$ T, measured at $D = -0.69$ V/nm and $v = 1.01$. Solid and dashed lines indicate different $B_\parallel$ sweeping directions. (**K**) Landau fan diagrams of $R_{yx}$ under $B_\parallel = 0.8$ T measured at $D = -0.69$ V/nm. White dashed lines show the expected evolution of Chern number $C = \pm 1$ QAH states based on the Streda formula. (**L**) Out-of-plane magnetic hysteresis loops measured at $D = -0.69$ V/nm, $v = 0.98$ and $B_\parallel = 0.1$ T, under different in-plane rotation angles $\theta_B = 0°-180°$. The inset shows the dependence of the out-of-plane magnetic hysteresis window $\Delta B_C$ on $\theta_B$. (**M**) Out-of-plane magnetic hysteresis loops measured under different $B_\parallel$ at $D = -0.69$ V/nm and $v = 0.98$. For small $B_\parallel = 0$ T and 0.1 T, the corresponding $B_C$ are ~ 0.04 T and ~ 0.14 T, respectively. As the $B_\parallel$ is further increased, the chirality

of the QAH state reverses. For larger $B_\parallel$ = 0.6 T and 0.9 T, the measured $B_C$ are ~ 0.4 T and ~ 0.25 T, respectively.